\documentclass[twocolumn]{aastex631}

\usepackage{epsfig}
\usepackage{xcolor}
\usepackage{amssymb}
\usepackage{pifont}

\usepackage{romannum}
\usepackage{lipsum}
\usepackage{amsmath}
\usepackage{rotating}
\usepackage[referable]{threeparttablex}
\usepackage{afterpage}
\usepackage{float}

\received{}
\revised{}
\accepted{}
\submitjournal{ApJ}

\shorttitle{ Swift and LCO reverberation mapping of an SMBHB candidate }
\shortauthors{T.~Liu et al.}

\begin{document}

\title{ Intensive Swift and LCO monitoring of PG~1302$-$102: AGN disk reverberation mapping of a supermassive black hole binary candidate }

\correspondingauthor{Tingting Liu}
\email{tingting.liu@nanograv.org}

\author[0000-0001-5766-4287]{Tingting Liu}
\affiliation{Department of Physics and Astronomy, West Virginia University, P.O. Box 6315, Morgantown, WV 26506, USA}

\author[0000-0001-8598-1482]{Rick Edelson}
\affiliation{Department of Astronomy, University of Maryland, College Park, MD 20742-2421, USA}

\author[0000-0002-6733-5556]{Juan V. Hern\'{a}ndez Santisteban}
\affiliation{SUPA Physics and Astronomy, University of St Andrews, Scotland KY16 9SS, UK}

\author{Erin Kara}
\affiliation{MIT Kavli Institute for Astrophysics and Space Research, Cambridge, MA 02139, USA}

\author{John Montano}
\affiliation{Department of Physics and Astronomy, 4129 Frederick Reines Hall, University of California, Irvine, CA, 92697-4575, USA}

\author[0000-0001-9092-8619]{Jonathan Gelbord}
\affiliation{Spectral Sciences Inc., 30 Fourth Ave. Suite 2, Burlington MA 01803, USA}

\author[0000-0003-1728-0304]{Keith Horne}
\affiliation{SUPA Physics and Astronomy, University of St Andrews, Scotland KY16 9SS, UK}

\author[0000-0002-3026-0562]{Aaron J. Barth}
\affiliation{Department of Physics and Astronomy, 4129 Frederick Reines Hall, University of California, Irvine, CA, 92697-4575, USA}

\author[0000-0002-8294-9281]{Edward M. Cackett}
\affiliation{Department of Physics and Astronomy, Wayne State University, 666 W. Hancock St, Detroit, MI 48201, USA}

\author[0000-0001-6295-2881]{David L.\ Kaplan}
\affiliation{Center for Gravitation, Cosmology and Astrophysics, Department of Physics, University of Wisconsin-Milwaukee, Milwaukee, WI 53211, USA}


\begin{abstract}
We present an intensive multiwavelength monitoring campaign of the quasar PG~1302$-$102 with Swift and the Las Cumbres Observatory network telescopes. At $z\sim0.3$, it tests the limits of the reverberation mapping (RM) technique in probing the accretion disk around a supermassive black hole (SMBH) and extends the parameter space to high masses and high accretion rates. This is also the first time the RM technique has been applied to test disk structures predicted in the SMBH binary model that has been suggested for this source. PG~1302$-$102 was observed at a $\sim$daily cadence for $\sim 9$ months in 14 bands spanning from X-ray to UV and optical wavelengths, and it shows moderate to significant levels of variability correlated between wavelengths. We measure the inter-band time lags which are consistent with a $\tau \propto \lambda^{4/3}$ relation as expected from standard disk reprocessing, albeit with large errors. The disk size implied by the lag spectrum is consistent with the expected disk size for its black hole mass within uncertainties. While the source resembles other reverberation-mapped AGN in many respects, and we do not find evidence supporting the prevalent hypothesis that it hosts an SMBH binary, we demonstrate the feasibility of studying SMBH binaries from this novel angle and suggest possibilities for the LSST Deep Drilling Fields. 
\end{abstract}


\section{Introduction} \label{sec:intro}

Understanding accretion onto supermassive black holes (SMBHs) in the centers of active galactic nuclei (AGN) and probing the innermost regions of AGN are areas of considerable interest in modern astrophysics. Due to its small physical size ($\sim$hundreds--thousands gravitational radii), the vicinity of the central engine cannot be directly imaged or resolved (with rare exceptions such as the observations of Sgr A$^{\star}$ and M87$^{\star}$ by the Event Horizon Telescope; e.g., \citealt{EHT2019,EHT2022}), but it is possible to utilize the temporally variable nature of an AGN to indirectly probe the size and geometry of the system.

One such indirect technique is known as reverberation mapping (RM; e.g., \citealt{Blandford1982,Peterson1993}; see \citealt{Cackett2021} for a recent review). The basic idea is that different emitting regions respond to the illuminating high-energy band, so that variations at different wavelengths are correlated but offset by a time lag which is determined by the light travel time to the emitting region. Therefore, with a temporal resolution of $\sim $ day, this technique can resolve a physical scale of $\sim$ light-days --- that of an AGN accretion disk. Several studies have performed intensive monitoring of AGN in multiple wavebands simultaneously (e.g., \citealt{Shappee2014,McHardy2014,McHardy2016,McHardy2018,Edelson2015,Edelson2017,Edelson2019,Cackett2018,Cackett2020,Hernandez2020,Vincentelli2021,Vincentelli2022, Kara2021,Kara2023}) and provided critical observational tests of predictions of the standard AGN disk and the reprocessing models. These studies have generally confirmed (1) the strong correlation between UV and optical bands (and, to a lesser degree, the X-ray band), with the longer wavelengths lagging behind shorter wavelengths (at least on short timescales); (2) the wavelength-dependent lags largely follow the $\tau \propto \lambda^{4/3}$ relation, which is the prediction of reprocessing from a standard disk \citep{Cackett2007}. The normalization of this relation informs the scale of the disk, which can be compared with the predicted disk size given the BH mass and mass accretion rate.

The AGN so far targeted by intensive broadband reverberation mapping (IBRM) campaigns are nearby ($z\lesssim0.1$), have BH masses between $\sim10^{7}-10^{8} M_{\odot}$, and mostly accrete at $\sim$ a few percent of the Eddington limit. In this work, we extend the IBRM experiment to PG~1302$-$102 (hereafter PG 1302), a quasar at $z = 0.2784$, which has a BH mass as high as $\sim 10^{9.4} M_{\odot}$ and appears to be accreting approximately at the Eddington limit (e.g., \citealt{Graham2015Nat}). The source is well known in the AGN literature as it is a bright ($V \sim$ 15 mag) and nearby quasar but has garnered significant additional interest in recent years: it has been suggested that it is the possible host of an SMBH binary (SMBHB), based on evidence of a large-amplitude ($\sim0.14$ mag), almost sinusoidal variation in its long-term $V$-band light curve from the Catalina Real-time Transient Survey (\citealt{Graham2015Nat}). The leading interpretation of this putative periodic variability is the relativistic Doppler boosting of emission mainly from the accretion disk (the so-called ``minidisk'') attached to the less massive, secondary BH \citep{D'Orazio2015}; hence the line-of-sight velocity of the black hole is imprinted on the light curve as a periodic variation on the orbital timescale ($\sim 5$ years in the observed frame). This picture is partly supported by numerical simulations, which reveal that in an unequal-mass SMBHB system, gas is preferentially accreted onto the secondary BH \citep{Farris2014,Duffell2020}, forming a smaller, but brighter, minidisk which dominates emission in the UV/optical band. Indeed, the variability amplitude of the source would imply a mass ratio of $q\sim0.1$, and the black hole mass must be larger than $\sim 10^{9} M_{\odot}$ in order to reproduce the observed large-amplitude variation. This mass ratio suggests a secondary minidisk size of $\sim 0.2a$ \citep{Artymowicz1994,Farris2014}, where $a$ is the binary orbital separation (which would be $\sim$ 0.01\,pc for this source).

The definitive detection of SMBHBs would be of fundamental importance. They are the expected outcome of galaxy mergers and structure formation (e.g., \citealt{Begelman1980}), but have remained observationally elusive despite significant search efforts (see, e.g., \citealt{Bogdanovic2022} for a recent review). The primary challenge is the small separation of the BHs, which cannot be directly resolved by current observatories (with rare exceptions; \citealt{Rodriguez2006}), leaving indirect observations the main and --- until their individual detections via gravitational waves\footnote{There is now evidence for a stochastic background of nanohertz gravitational waves from pulsar timing array experiments (e.g., \citealt{NG15yrGWB}). The amplitude and spectral shape of this background are consistent with expectations from a population of SMBHBs (e.g., \citealt{NG15yrAstro}). The Laser Interferometer Space Antenna \citep{Amaro-Seoane2017} will be sensitive to merging massive black hole binaries in the millihertz frequency range.} --- the only approach. However, the binary nature of PG 1302 is still a matter of debate, and much of the debate focuses on whether the variability of the source is indeed periodic. It has been noted that the red noise characteristic of normal AGN variability could mimic periodicity over a few short cycles with irregular sampling and large photometric uncertainties \citep{Vaughan2016}, and longer observations will show that the ``periodicity'' will no longer stay persistent \citep{Liu2018}. Therefore, the confirmation of PG 1302's binarity should be supported by evidence for significant departures from a normal AGN in aspects other than its (putative) periodicity, and the absence of such evidence should be interpreted as evidence against the binary hypothesis.

In this work, we apply the IBRM technique to search for such evidence in PG~1302, with the objective of testing for the presence of a minidisk. PG 1302 is one of the few known cases where such experiments may be possible, because while minidisks can form in SMBHB systems, they may not be an ubiquitous feature in all SMBHBs, and interpretations of the observations of candidate systems do not always require the existence of minidisks. For instance, while the well-known SMBHB candidate OJ 287 (see e.g., \citealt{Valtonen2021} for a recent review) has been the target of intensive multi-wavelength monitoring (e.g., \citealt{Komossa2021MOMO}), its binary model invokes the secondary BH impacting the accretion disk of the primary BH. As the length of the observing campaign is short relative to the reported orbital timescale, our experiment is agnostic to the periodicity (or not) of the source, and instead directly probes disk features that are key predictions of its binary model.


\begin{table*}[ht]
\caption{Summary of Observations}
\begin{center}
\begin{tabular}{lcccc}
\hline \hline
 \hline
Filter & $\lambda_{\rm eff}$ (\AA) & MJD range & Number of epochs  & $\Delta t$ (day$^{-1}$) \\ 
\hline
Swift & &  &  &  \\ 
X-ray & \phn\phn\phn6 & 59551--59816 & 268 &  0.6, 0.7, 1.1  \\ 
$W2$ & 1928 & 59551--59816 & 210 & 0.9, 0.7, 1.3  \\ 
$M2$ & 2246 & 59552--59816 & 196 & 1.0, 0.8, 1.4    \\ 
$W1$ & 2600 & 59552--59816 & 206 & 0.9, 0.8, 1.4   \\ 
$U$ & 3465 & 59551--59816 &  221 & 0.8, 0.7, 1.3    \\ 
$B$ & 4392 & 59551--59815 & 231  & 0.6, 0.8, 1.3   \\ 
$V$ & 5468 & 59551--59816  & 243 & 0.7, 0.7, 1.2    \\ 
\hline
LCO &  &  &  &    \\ 
$u$ & 3580 & 59561--59817 & 153 &  0.9  \\ 
$B$ & 4392 & 59561--59817 & 166 &  0.8  \\ 
$g$ & 4770 & 59561--59817 & 174 &  0.8  \\ 
$V$ & 5468 & 59561--59817 & 178 &  0.8  \\ 
$r$ & 6215 & 59561--59817 & 169 &  0.8   \\ 
$i$ & 7545 & 59561--59817 & 164 &  0.8  \\ 
$z$ & 8700 & 59561--59817 & 116 &  1.2  \\ 
\hline \hline
\multicolumn{5}{l}{
\begin{minipage}{3.5in}~\\
Note: Consecutive observations at the same LCO telescope on the same day are considered one epoch. The mean sampling rate $\Delta t$ is estimated for the period MJD 59551--59598, 59628--59670, 59683--59817 (Swift; the target was not observed during the two long gaps) and MJD 59650--59750 (LCO). LCO telescope \texttt{1m004} was removed from the intercalibration procedure due to photometry/calibration issues.
\end{minipage}
}\\
\end{tabular}
\end{center}
\label{tab:obs}
\end{table*}


The paper is organized as follows: we describe our observing campaign and the multi-band data in \S \ref{sec:obs}, and in \S \ref{sec:analysis}, we measure the cross-correlations and inter-band lags between the light curves. In \S \ref{sec:results} we compare the lag measurements with predictions of a standard AGN disk as well as the binary model. We summarize the results in \S \ref{sec:conclude}.


\section{Observations} \label{sec:obs}

\subsection{Swift}\label{sec:swift}

The Neil Gehrels Swift Observatory \citep{Gehrels2004} observed PG 1302 from 2021 December 3 to 2022 August 25 at a $\sim$ half-day cadence for the first $\sim 100$ days and a $\sim$ daily cadence for the remainder of the campaign (Swift Cycle 17; proposal number 1720044; PI: T. Liu). The X-ray Telescope (XRT; \citealt{Burrows2005}) observations were carried out in photon counting mode, and Ultraviolet/Optical Telescope (UVOT; \citealt{Roming2005}) observations were made in six-filter image mode (\texttt{0x30ed} for the first half of the campaign and \texttt{0x224c} for the second half) in the $W2, M2, W1, U, B, V$ filters. In 2022~January, the spacecraft experienced a reaction wheel anomaly; this resulted in a 30-day gap in our observing campaign. A second, 13-day gap occurred when the source was close to the anti-Solar direction and could not be scheduled due to the spacecraft's slewing constraints. The final Swift campaign spanned nine months with $\sim 300$ observations in each filter.

The XRT light curve was generated using the data analysis tool\footnote{\url{http://www.swift.ac.uk/user\_objects}} \citep{Evans2009} with snapshot binning. The UVOT data were processed following the same procedures as \cite{Hernandez2020}, including the removal of dropout flux points. As the source occasionally falls within regions of reduced sensitivity on the detector, the UVOT light curves exhibit a number of unphysical, low flux points. These points are screened and only data which survive the ``Low'' sensitivity mask are used in the analysis. We refer the reader to previous work (e.g., \citealt{Hernandez2020}) for details of the reduction and screening procedures. We summarize the final Swift dataset in Table \ref{tab:obs} and present the light curves in Figure \ref{fig:lc} and Table \ref{tab:sw_lc}. 


\begin{table}[ht]
\caption{ Swift UV and optical light curves}
\begin{center}
\begin{tabular}{lccc}
\hline
Filter & MJD & Flux & Flux error \\ 
&  & \multicolumn{2}{c}{(10$^{-15}$\,erg\,cm$^{-2}$\,s$^{-1}$\,\AA$^{-1}$)}  \\
\hline
$W2$ &  59551.051 & 18.045 & 0.542 \\ 
$W2$ & 59552.887 & 17.430 & 0.289 \\
\nodata & \nodata & \nodata & \nodata \\
$M2$ & 59552.891 &15.724 & 0.315 \\
$M2$ & 59553.622 & 15.668 & 0.332 \\
\nodata & \nodata & \nodata & \nodata \\
$W1$ & 59552.883 & 13.117 & 0.271 \\
$W1$ & 59553.614 & 13.051 & 0.264\\
\nodata & \nodata & \nodata & \nodata \\
$U$ & 59551.691 & \phn9.738 & 0.217 \\
$U$ & 59552.884 & 10.010 & 0.227 \\
\nodata & \nodata & \nodata & \nodata \\
$B$ & 59551.692 &  5.915 & 0.155 \\
$B$ & 59552.885 & 5.928 & 0.160 \\
\nodata & \nodata & \nodata & \nodata \\
$V$ & 59551.697 & 3.877 & 0.163 \\
$V$ & 59553.039 & 3.721 & 0.168 \\
\hline \hline
\multicolumn{4}{l}{
\begin{minipage}{2in}~\\
Note: The full table is available in a machine-readable format.
\end{minipage}
}
\end{tabular}
\end{center}
\label{tab:sw_lc}
\end{table}


\begin{table}[ht]
\caption{Inter-calibrated LCO light curves}
\begin{center}
\begin{tabular}{lccc}
\hline
Filter & MJD & Flux & Flux error \\ 
&  & (mJy) & (mJy) \\
\hline
$u$ &  59257.354 &  3.581 &  0.024 \\ 
$u$ & 59259.374 & 3.585 & 0.024 \\
\nodata & \nodata & \nodata & \nodata \\
$B$ & 59418.023  & 3.286 & 0.050 \\ 
$B$ & 59430.015& 3.272 & 0.039 \\
\nodata & \nodata & \nodata & \nodata \\
$g$ & 59256.329  & 3.531 & 0.038 \\ 
$g$ & 59257.360 & 3.536 & 0.022 \\
\nodata & \nodata & \nodata & \nodata \\
$V$ & 59418.026  & 3.480 & 0.029 \\ 
$V$ & 59424.408 & 3.465 & 0.018 \\
\nodata & \nodata & \nodata & \nodata \\
$r$ & 59256.332  & 3.797 & 0.019 \\ 
$r$ & 59257.362 & 3.795 & 0.023 \\
\nodata & \nodata & \nodata & \nodata \\
$i$ &  59256.334 & 3.903 & 0.028 \\ 
$i$ & 59257.365 & 3.946 & 0.023 \\
\nodata & \nodata & \nodata & \nodata \\
$z$ & 59256.338  & 5.211 & 0.122 \\ 
$z$ & 59257.369 & 5.129 & 0.098 \\
\hline \hline
\multicolumn{4}{l}{
\begin{minipage}{2in}~\\
Note: The full table is available in a machine-readable format.
\end{minipage}
}
\end{tabular}
\end{center}
\label{tab:lco_lc}
\end{table}


\subsection{Las Cumbres Observatory}\label{sec:lco}

The Swift observations were coordinated with intensive ground-based monitoring with the Las Cumbres Observatory Global Telescope (LCO; \citealt{Brown2013}) network in seven optical bands: SDSS $u g r i$, Pan-STARRS $z$, and Johnson $B$, $V$. LCO observed PG 1302 from 2021 February 11 to 2021 August 20, 2021 December 13 to 2022 August 26, and again from 2022 December 2 and 2023 August 16 at a $\sim$daily to few-day cadence, and a total of twelve 1-m robotic telescopes participated in the campaign (as part of the LCO Key Project 2020B-006; PI: J. V. Hern\'{a}ndez Santisteban). The images were processed with the \texttt{BANZAI} pipeline \citep{McCully2018}, and we refer the reader to \cite{Hernandez2020} for a detailed description of the data reduction and photometry procedure.

\begin{figure*}[ht]
\centering
\epsfig{file=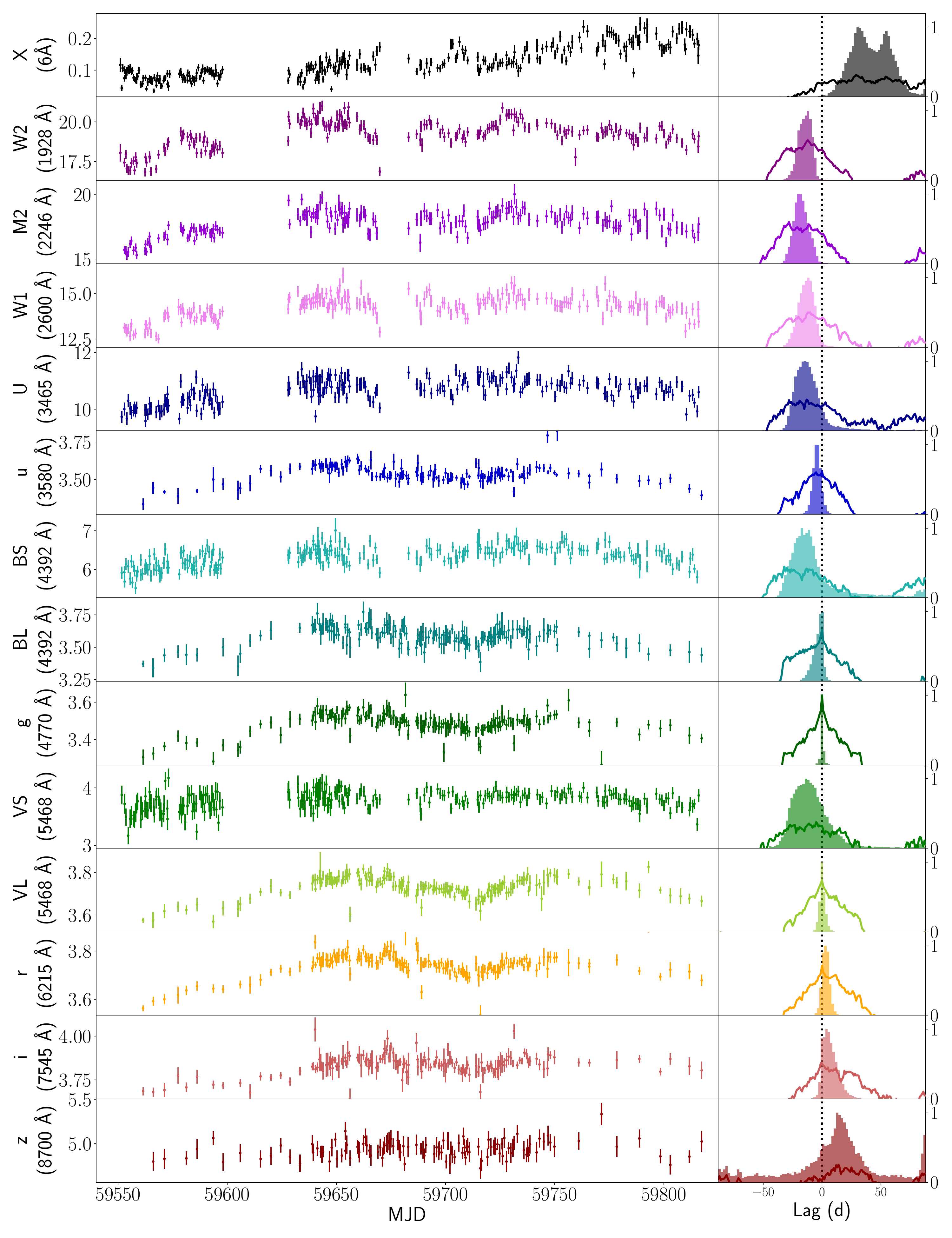,width=0.9\textwidth,clip=}
\caption{Multi-band light curves of PG~1302$-$102 from the observing campaign in the X-ray (first panel), UV (next three panels), and optical (remaining panels; for $B$ and $V$ bands, ``S'' denotes Swift and ``L'' denotes LCO). The Swift XRT data are in units of counts\,s$^{-1}$, and the UV/optical data are in units of 10$^{-15}$\,erg\,cm$^{-2}$\,s$^{-1}$\,\AA$^{-1}$. The LCO data are in units of mJy. The right panels are the CCFs and centroid histograms in the respective bands relative to the reference band $g$. The zero lag is marked with black dotted lines to guide the eye. }
\label{fig:lc}
\end{figure*}

We then follow a few additional steps to refine the light curves: since each visit at a given telescope typically consisted of two consecutive exposures, we first combine the fluxes using a weighted average to create a single observation. We then inter-calibrate the LCO light curves in the same band from different telescopes, since it has been noted by \cite{Hernandez2020} that systematic flux offsets exist even between telescopes with identical designs and filters. To do so, we use the package \texttt{PyROA}\footnote{\url{https://github.com/FergusDonnan/PyROA}} \citep{Donnan2023}, which models the light curve as a running optimal average (ROA) with a scale factor $A$ and a flux offset $B$ for each telescope. The algorithm then produces a merged light curve by optimizing these two parameters (where $\langle$A$\rangle$$\approx$1 and $\langle$B$\rangle$$\approx$0) using Markov Chain Monte Carlo (MCMC). The model also includes an extra noise parameter for each telescope to account for the fact the reported flux uncertainties are underestimated (see \citealt{Hernandez2020}); this parameter is then added in quadrature to the nominal uncertainties. The flexibility of the ROA is determined by a window width parameter, and thus more flexible models are able to capture more rapid variations. Outliers are determined by comparing them to the ROA model generated to merge the light curves: those observations that remain above the chosen $4\sigma$ threshold are forced to remain at that level by expanding their error bars accordingly. This procedure allows for down-weighting their importance to the model, without removing the data points. After the inter-calibration procedure, points with significantly larger error bars, including those produced during the aforementioned ``soft'' outlier sigma clipping process by \texttt{PyROA}, are removed. All data collected between 2021 February and 2023 August are used in the intercalibration procedure, and the light curves used in the cross-correlation analysis are from the second monitoring period (which was simultaneous with the Swift campaign). They are shown in Figure \ref{fig:lc}, and a summary of observations is provided in Table \ref{tab:obs}. The light curves from the full monitoring period are presented in Table \ref{tab:lco_lc}.  

Additionally, the 0.5-m robotic telescope at the Dan Zowada Memorial Observatory \citep{Carr2022} also observed the source from 2021 February 9 to 2022 June 6 in $g r i z$ bands for $\sim40$ epochs in each band. However, due to their larger measurement uncertainties, the Zowada data are not used in the cross-correlation analysis.


\begin{table*}
\caption{Interband lags}
\begin{center}
\begin{tabular}{l c ccc c}
\hline
 & & \multicolumn{3}{c}{ICCF} & \multicolumn{1}{c}{\texttt{PyROA}}\\
 \hline
Filter  & $f_{\rm var}$ &  $\tau_{\rm median}$ & Uncertainty & $r_{\rm max}$ & $\tau$ \\ 
& & (days)  & (days)  & & (days)  \\
\hline
Swift & & \\
X-ray  & 0.381& \phn\phd40.86 & \phn\phn(25.39, 58.55)   & 0.31 & \nodata \\ 
$W2$  & 0.043 &$-$14.14  & ($-$20.13, $-$8.97)  & 0.57 &  $-$7.88$^{+1.40}_{-1.51}$  \\
$M2$  & 0.044 & $-$17.86   &  ($-$23.56, $-$10.98) & 0.60 &  $-$7.90$^{+1.70}_{-1.43}$ \\ 
$W1$ & 0.037  & $-$13.09  &  ($-$20.15, $-$6.71) & 0.50 & $-$6.58$^{+1.42}_{-1.62}$ \\ 
$U$  & 0.033  &$-$12.95  &  ($-$21.69, $-$2.00) & 0.45 & $-$8.79$^{+1.90}_{-2.09}$ \\
$B$ & 0.032  & $-$12.50  & \phn\phn($-$23.86, 9.27)  & 0.45 &  $-$9.08$^{+2.03}_{-1.99}$  \\
$V$  & 0.021  &$-$11.04  &  \phn\phn($-$23.74, 4.95) & 0.38 &  $-$10.01$^{+2.45}_{-2.32}$  \\
\hline
LCO & & \\
$u$ & 0.014  & $-$4.26  & ($-$7.70, $-$0.90)  & 0.62 &  $-$6.70$^{+1.15}_{-1.10}$ \\  
$B$  & 0.013  &$-$2.50  &  \phn($-$7.90, 0.01) & 0.74 &  $-$1.68$^{+1.40}_{-1.34}$ \\ 
$g$  & 0.013  & \phn\phn\phn\phn0  & \phn\phd($-$0.86, 0.92)   & 1.00 & 0  \\ 
$V$ & 0.010  & \phn\phd0.01  &  \phn\phd($-$1.91, 2.65) & 0.74  &   \phn2.20$^{+1.11}_{-1.18}$  \\ 
$r$ & 0.011  &\phn\phd3.14  & \phn\phn\phn(0.00, 6.15)  & 0.71 &   \phn4.05$^{+1.24}_{-1.17}$  \\ 
$i$  & 0.011  & \phn\phd6.05  &  \phn\phn(0.69, 13.51) & 0.51  &   \phn8.54$^{+1.59}_{-1.59}$   \\ 
$z$ & 0 &  \phn13.30  & ($-$28.06, 37.96)  & 0.25  &  \phn5.47$^{+6.03}_{-5.65}$  \\ 
\hline \hline

\multicolumn{6}{l}{
\begin{minipage}{4in}~\\
Note: $\tau_{\rm median}$ is the median of the centroid distribution, and the uncertainty is given by the middle 68\% of the distribution. A positive value indicates that the band lags behind the reference band. $r_{\rm max}$ is the peak of the CCF. $f_{\rm var}$ is the fractional variability amplitude: $f_{\rm var}$=$\sqrt{\frac{S^{2}-\overline{\sigma^{2}}}{\bar{x}^{2}}}$ (e.g. \citealt{Vaughan2003}), which quantifies the level of intrinsic variability. (The uncertainties of the $f_{\rm var}$ are negligible and therefore not shown.)
Note that $f_{\rm var}$ in the $z$ band is forced to be zero because the term inside the square root is negative. The photometric uncertainties in this band are likely overestimated due to fringing.
\end{minipage}
}
\end{tabular}
\end{center}
\label{tab:lags}
\end{table*}


\section{Time series analysis}\label{sec:analysis}

\subsection{Cross-correlation functions}\label{sec:CCF}

We use the \texttt{R} package \texttt{sour}\footnote{\url{https://github.com/svdataman/sour}} to compute the cross-correlation between the Swift XRT, UV, and LCO optical bands, initially using Swift $W2$ as the reference band as in previous IBRM campaign analyses. However, there appear to be discrepancies in the lag measurements at the $> 1 \sigma$ (but $<2 \sigma$) level in similar wavebands (e.g., $U$ and $u$). After ruling out any issues due to LCO photometry or inter-calibration (we have remeasured the LCO light curves using an independent photometry pipeline that follows the method described in \citealt{Kara2021} and a different intercalibration procedure \texttt{PyCALI}; \citealt{Li2014}), we speculate that the differences are due to the different sampling patterns of the two observatories. Hence, when the cross-correlation is measured relative to $W2$ (which suffers from long gaps), they cause the ICCF to respond differently in similar wavebands (one of which shares the same gaps while the other does not). Therefore, to mitigate the biases that are potentially introduced to the lag measurements, we adopt the LCO $g$ band (which has more even sampling), as the reference band in the analysis presented in this section.


\begin{figure}[t]
\centering
\epsfig{file=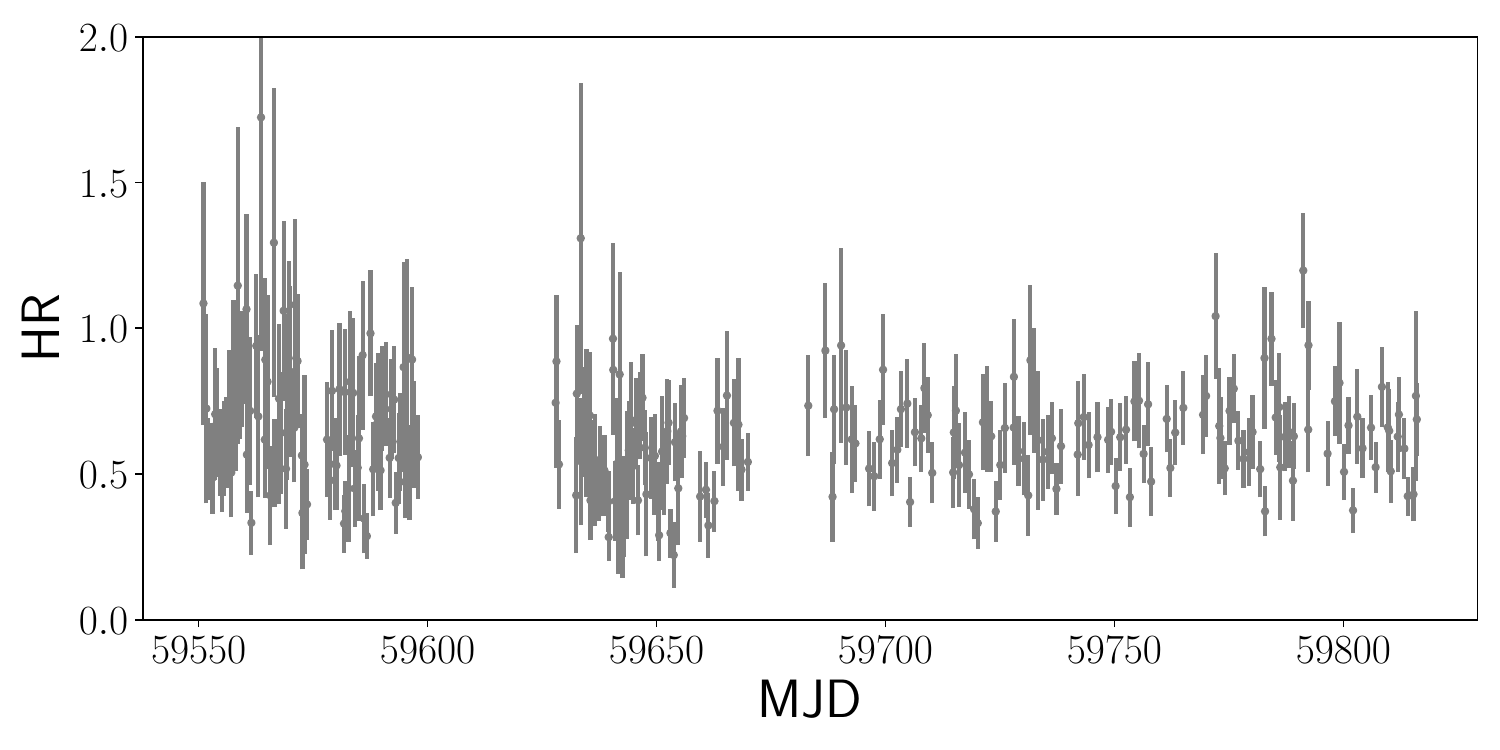,width=0.45\textwidth,clip=}
\epsfig{file=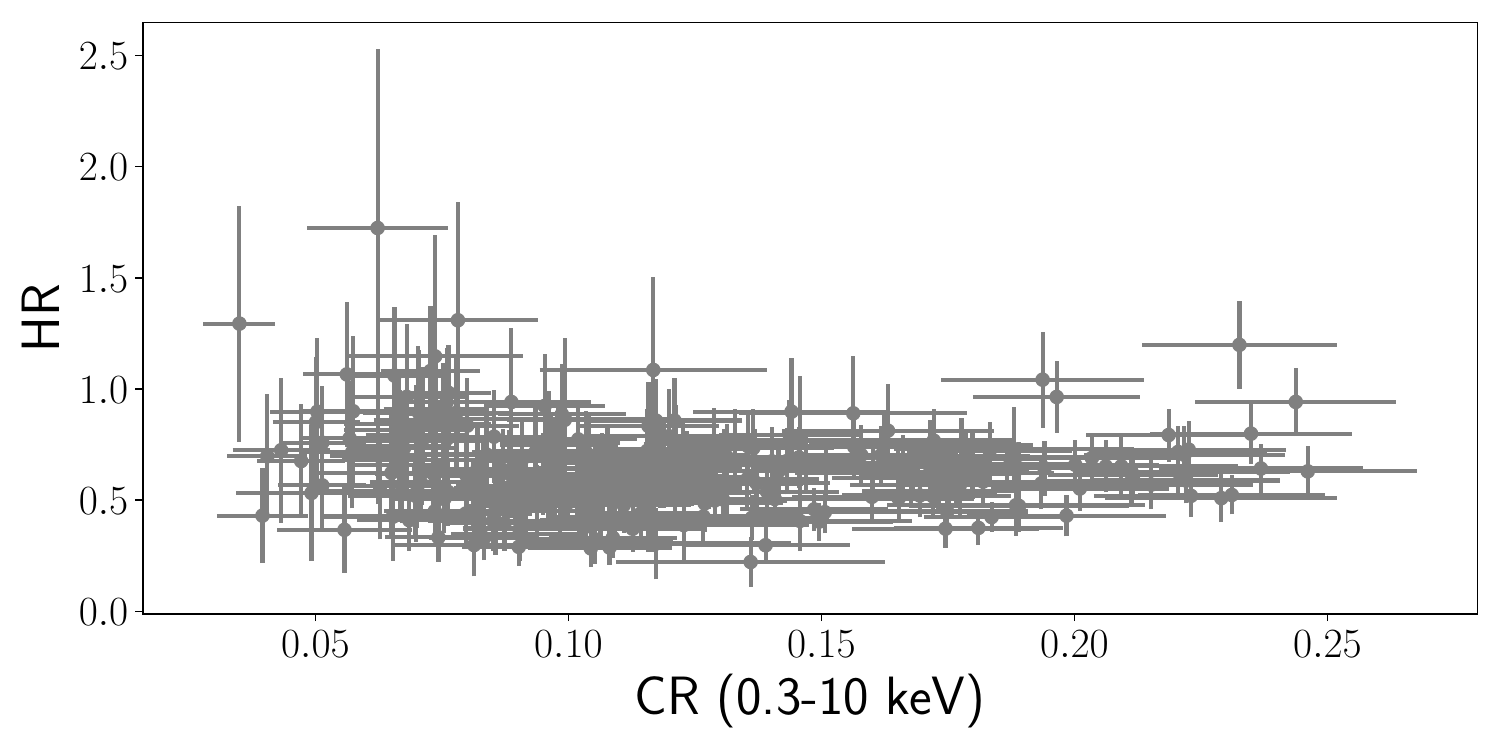,width=0.45\textwidth,clip=}
\caption{ The hardness ratio (HR; see text for the definition) time series (upper panel) and the HR versus the count rate (CR; lower panel). The HR remains largely constant with no discernible trends as a function of time or CR. }
\label{fig:hr}
\end{figure} 


For each pair of light curves, we first compute the interpolated cross-correlation function (ICCF; e.g., \citealt{Gaskell1986}) using two-way interpolation (so that the autocorrelation function is symmetric). We use a lag bin width of $\Delta \tau=1$ day, and a maximum lag at which to evaluate the ICCF that equals 1/3 of the observation duration: $\texttt{max.lag}$=88 days. Next, we apply the flux randomization and random subset sampling method (FR/RSS; e.g., \citealt{Peterson1998}) to obtain the uncertainty on the measured lag. The RSS process randomly selects a subset of the parent light curve, and the FR process modifies the flux measurements by random Gaussian deviates which are based on the reported errors. Thus, in the resulting realization, both the fluxes and sampling times are randomized. This procedure is repeated for 20000 realizations and produces a distribution of CCF centroids which are calculated using CCF points higher than 80\% of the peak value.

We tabulate the lags and peak CCF correlation coefficients in Table \ref{tab:lags}. The cross-correlation analysis shows that most UV and optical bands are well-correlated with the LCO $g$ band with well-constrained (positive and negative) lags. By contrast, the X-ray light curve is very poorly correlated with the UV and optical bands ($r_{\rm max}=0.31$). This can also be seen via visual inspection of the light curves: the source appears to brighten overall in the X-ray during the campaign, but this trend is not mirrored in the UV and optical bands. Changes in obscuration could cause the UV/optical variability in an AGN to become disconnected from the X-ray variability, and these changes would manifest as variations in the hardness ratio (defined here as HR=H/S, where H and S are the 1.5-10 keV and 0.3-1.5 keV count rates, respectively). Yet there are no significant changes in the HR (fractional variability amplitude of the time series = 0.06$\pm$0.07) or correlations between the HR and the X-ray count rate (Spearman's rank p-value = 0.33; Figure \ref{fig:hr}). Further, the X-ray band appears to lag behind UV/optical, which is not expected from the canonical picture of an X-ray emitting corona illuminating the disk and driving UV/optical variability. Nor is the long lag ($\sim 46$ days) consistent with a simple light travel time effect with the corona located at a low height above the disk. Opposite X-ray lags have been previously observed in at least some AGN (e.g., \citealt{Edelson2019, Kara2023}). A possible explanation could be that a fluctuation propagates inward on a timescale longer than the light travel time (e.g. \citealt{Arevalo2008}) and therefore appears later in the X-ray. However, because of the weak cross-correlation of the X-ray band, this opposite lag is highly uncertain.

\subsection{Comparisons with \texttt{PyROA} lag measurements}\label{pyroa}

We have additionally utilized the lag fitting functionality of the \texttt{PyROA} package \citep{Donnan2021} to measure the inter-band lags. Compared to the ICCF, \texttt{PyROA} is able to better handle large observing gaps in the light curves, because it gathers variability information from all bands and interpolates across gaps with error envelopes. The algorithm works similarly to the inter-calibration process, by first scaling and offsetting the multi-band light curves. It further shifts the light curves by the time lag parameters and effectively stacks them to create a merged time series, for which a ROA model can then be determined. The parameter estimates and uncertainties are obtained from the MCMC samples.

We compute the lags relative to the LCO $g$ band, and report the measurements in Table \ref{tab:lags}. \texttt{PyROA} have detected lags that increase overall with wavelength, which is consistent with the ICCF results. Note that \texttt{PyROA} produces much smaller uncertainties than the ICCF, which is in agreement with the comparison made in \cite{Donnan2021}. Nonetheless, most \texttt{PyROA} measurements are consistent with the ICCF-measured lags within one-sigma uncertainties.


\section{Results}\label{sec:results}

\subsection{Interband lag fits}\label{sec:fits}

\begin{figure}[ht!]
\centering
\epsfig{file=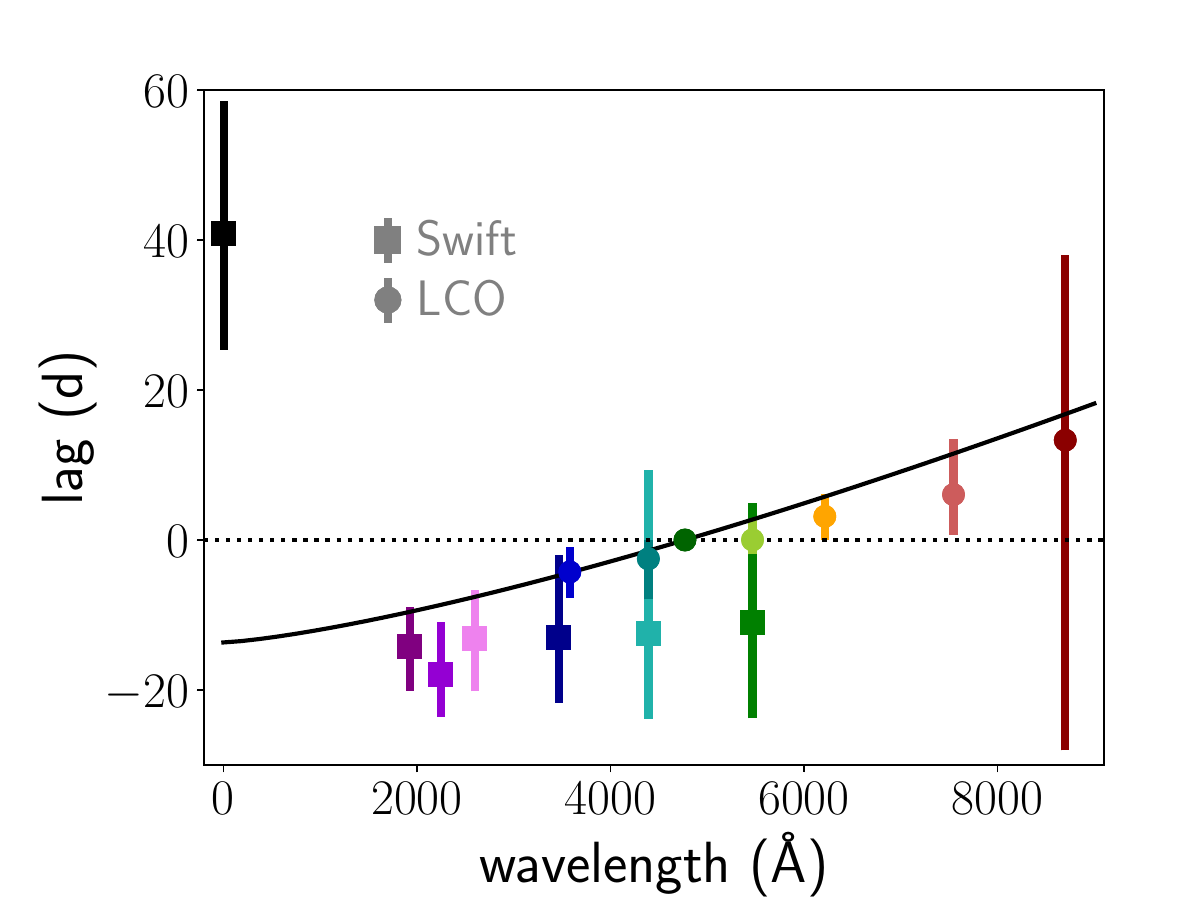,width=0.47\textwidth,clip=}
\caption{The lag spectrum fitted with a $\tau \propto \lambda^{4/3}$ relation as predicted for a standard disk.
The X-ray lag (black square) was not used in the fit. The normalization $\tau_{0}$ of this relation is determined by the y-intercept. The reference band is LCO $g$.
The color scheme is the same as in Figure \ref{fig:lc}. The wavelengths and lags are in the observed frame.
}
\label{fig:lags}
\end{figure}

In Figure \ref{fig:lags}, we show the inter-band lag spectrum, i.e., lags as a function of wavelength, measured by the ICCF using LCO $g$ as the reference band. It shows a trend of longer lags at longer wavelengths, consistent with previous results of AGN disk RM campaigns. To quantify this relation, we fit the lag spectrum (excluding the X-ray band which is poorly correlated with UV/optical with an opposite lag) with a function of the form $\tau = \tau_{0}[(\lambda/\lambda_{0})^{\beta}-1]$, where $\beta$=4/3 is the prediction of reprocessing in a standard disk \citep{Cackett2007}, and $\lambda_{0}$ corresponds to the reference $g$ band. The lag distributions are slightly asymmetric but are sufficiently close to being Gaussian (except for the $z$ band); therefore, we approximate the 1-$\sigma$ uncertainty by averaging the upper and lower errors. The Swift and LCO lags in similar bands are combined using a weighted average. We hence obtain $\tau_{0}=13.7 \pm 2.9$\,days\footnote{All $\tau_{0}$ values below are quoted in the observed frame.}. As can be seen from the lag spectrum, despite the large uncertainties, the UV/optical lags across all observed wavelengths are consistent with reprocessing in a standard disk.


\subsection{Comparing with expected disk size}\label{sec:interp}

We proceed to compare the measured disk size (represented by the normalization factor $\tau_{0}$ that corresponds to the reference band) with the expectations from a geometrically thin, optically thick disk associated with a black hole of its mass. We use the relation between radius and emitting wavelength $\lambda$ in \cite{Fausnaugh2016}: 

\[ \alpha=\frac{1}{c}\left(X\frac{k\lambda}{hc}\right)^{4/3} \left[\left(\frac{GM}{8\pi\sigma}\right)\left(\frac{L_{\rm Edd}}{\eta c^{2}}\right)(3+\kappa)\dot{m}_{\rm E}   \right]^{1/3}\, \textrm{days}, \] 

\noindent where $\dot{m}_{\rm E}$ is the Eddington ratio, $\eta$ is the radiative efficiency, we assume a negligible external to internal heating ratio ($\kappa=0$), and $X=4.87$ if we assume Wien's law when converting from temperature to wavelength at a given radius, and $X=2.49$ if a flux-weighted radius is assumed instead.


\begin{figure}[ht]
\centering
\epsfig{file=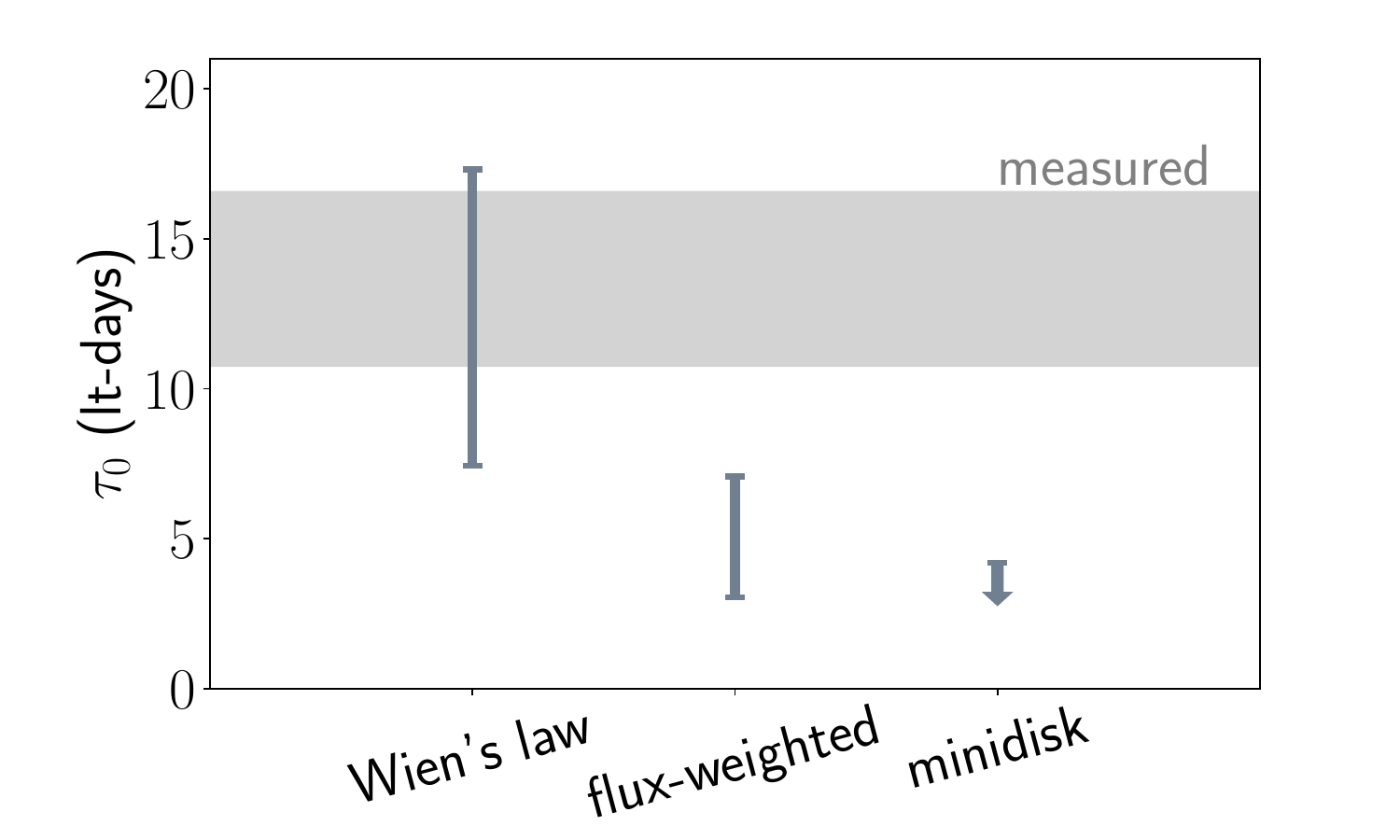,width=0.5\textwidth,clip=}
\caption{We interpret the measured disk size (represented by $\tau_{0}$) in three separate scenarios. The first two assume PG~1302$-$102 has a single black hole mass of $\log M_{\rm BH}=8.3-9.4$; the measured $\tau_{0}$ is consistent with the expected disk size within uncertainties assuming Wien's law but larger than the predicted size in the flux-weighted case. The third one considers the scenario where the observed disk is carried by the secondary in an SMBHB system (i.e., a minidisk). The measured disk size is larger than the expected minidisk size by at least a factor of $\sim$2. }
\label{fig:tau}
\end{figure}

PG 1302 has a large uncertainty on its black hole mass, ranging from $\log M_{BH}/M_{\odot} = $ 8.3 to 9.4. Assuming a steady bolometric luminosity $L_{\rm bol}=6.5\times10^{46}$ erg s$^{-1}$, this translates to a large uncertainty on the Eddington ratio: $\dot{m}_{\rm E}=0.2-2.6$. Here we adopt $\eta = 0.3$, which is suggested for luminous quasars \citep{Yu2002}. We therefore estimate that the expected $\tau_{0}$ is [7.4, 17.3] days for the Wien's law case, and [3.0, 7.1] days for the flux-weighted case. Since we measured $\tau_{0}\approx 14$ days from the lag spectrum fitting, this is larger than the predicted size in the flux-weighted case but consistent with Wien's law case within uncertainties (Figure \ref{fig:tau}).

We then consider the alternative interpretation that PG 1302 is a binary SMBH system. In this hypothesis, in order for the system to produce $V$-band periodicity at the observed amplitude, the total black hole mass must be at the higher end of the mass range, i.e., $\log M_{\rm BH} \approx 9.2 - 9.4$, and the mass ratio must be low, i.e., $q\approx 0.1$ \citep{D'Orazio2015}. As previously discussed in Section \ref{sec:intro}, in this low-mass ratio regime, mass is preferentially accreted onto the secondary BH by a factor of $\sim$10 (e.g., \citealt{Farris2014, Duffell2020}), and thus we consider the UV and optical bands to be dominated by emission from the secondary minidisk. Similarly, contamination from the outer, circumbinary disk is expected to be negligible, since the circumbinary disk would contribute to less than 20\% of the total luminosity in this system (see \citealt{D'Orazio2015}).


\begin{figure}[ht!]
\centering
\epsfig{file=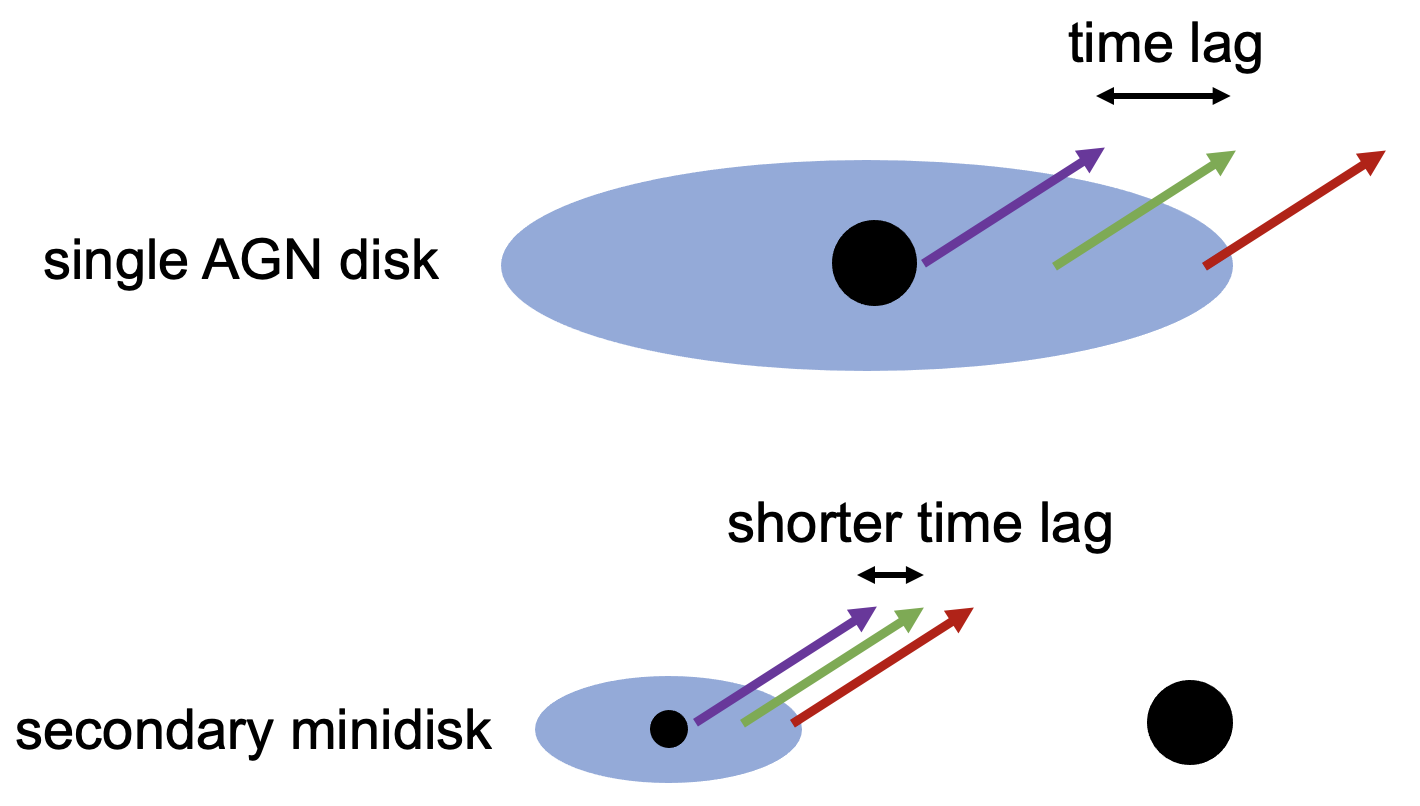,width=0.45\textwidth,clip=}
\caption{The measured inter-band lags infer the size of the accretion disk around the SMBH (upper figure) which can be independently estimated from its black hole mass and mass accretion rate. If PG~1302 were an SMBH binary, the minidisk around the secondary BH would result in shorter lags (lower figure).}
\label{fig:picture}
\end{figure}

We proceed to approximate the size of the secondary minidisk as the tidal truncation radius: $r_{\rm tid} \approx 0.19 a$ (for a $q=0.11$ binary; \citealt{Artymowicz1994,Farris2014}), where $a$ is the binary orbital separation which we can compute using the observed period of $P=1996$ days. We assume the edge of the minidisk emits in the $V$ band, and we consider this to be a conservative estimate because it is a necessary condition for the (putative) $V$-band periodicity. (More detailed modeling of a binary SED is possible; see, e.g., \citealt{Roedig2014,Guo2020}.) Hence, we estimate an \emph{upper limit} on the expected $V$-band time lag; in practice, this corresponds to a $\tau_{0}$ that has been extrapolated to the reference $g$ band assuming a $\lambda^{4/3}$ power law: $\tau_{0}$= [3.6, 4.2] lt-day. This is smaller than the measured $\tau_{0}$ by a factor of $\gtrsim 2$, and therefore we find the SMBHB model is in tension with the observed disk size (Figure \ref{fig:tau}; see Figure \ref{fig:picture} for a schematic representation). 

We note that these results should be interpreted with the caveat that they do not definitively rule out a smaller disk whose size has been overestimated due to the same mechanisms that cause other RM-measured AGN disk sizes to appear larger than predicted (typically by a factor of $\sim2-3$; e.g., \citealt{McHardy2014,Fausnaugh2016,Edelson2017,Cackett2018,Cackett2020}). A number of factors could have contributed to the difference between observed and predicted sizes (see, e.g., \citealt{Cackett2021} for a summary). For instance, uncertainties on the black hole mass and mass accretion rate affect the disk size, so does our limited knowledge of other parameters such as $\eta$ and the ratio of external to internal heating. Other possible physical reasons include departures from the standard disk model and contributions of the diffuse continuum emission (e.g., \citealt{Korista2019}) from the broad line region (which is at a larger physical distance and therefore lengthens the lags\footnote{We note, however, that we do not find clear evidence for longer timescales in the lag spectrum or the CCFs.}). Future IBRM studies will help elucidate this problem.


\section{Summary and Conclusions} \label{sec:conclude}

We have presented an intensive, multi-band monitoring campaign of the quasar PG~1302$-$102 over the course of $\sim 2.5$\,years. The campaign was coordinated between Swift and LCO during the second observing season, where the two observatories simultaneously monitored the source at a daily to sub-day cadence. It has the highest redshift and one of the highest accretion rates among the $\sim$ dozen of AGN that have been targeted by IBRM campaigns. It is also the first SMBHB candidate where the IBRM technique has been applied to test the predicted binary disk structure.

The source is variable on the observed timescales and shows correlations between UV and optical bands. The wavelength-dependent lag spectrum is consistent with the expected $\tau\propto\lambda^{4/3}$ relation for a standard AGN disk, consistent with other AGN observed by previous IBRM campaigns. The estimated disk size appears to be larger than the predicted size (also reminiscent of many other reverberation-mapped AGN) but can also be consistent with predictions within uncertainties, depending on parameter assumptions. Contributions from the diffuse BLR continuum could have lengthened the observed lags of a smaller accretion disk, however, they are not directly observed in the lag spectrum or the CCFs, and we are unable to investigate this possibility with the available data.

We further explore the possibility that PG 1302 hosts an SMBHB, which has been suggested based on evidence for a $\sim5$\,yr periodicity which is attributed to the relativistic beaming of a minidisk. However, we do not find evidence supporting the presence of such a minidisk. This conclusion is primarily based on the measured disk size being at least twice larger than that of the secondary SMBH in a binary system, for the sets of parameters that are required to produce the putative periodicity (although with the same caveat regarding BLR contributions). 

While PG 1302 represents a rare case where the IBRM technique can be applied to test an SMBHB hypothesis with current observatories, there may be opportunities to systematically search for SMBHBs in the Deep Drilling Fields (DDFs) of the Rubin Observatory Legacy Survey of Space and Time \citep{Ivezic2008}. The thousands of reverberation-mapped AGN in the DDFs \citep{Brandt2018} will have multi-band data similar to the LCO light curves presented here, in each observing season, and with better photometric precision. Assuming a two-day DDF cadence, SMBHB minidisks could manifest as smaller-than-expected or unresolved lags for a wide range of binary parameters. This would thus open another path towards finding and studying these rare objects. 

As we were finalizing the paper, \cite{Dotti2023} proposed that the (lack of) correlation between the red and blue parts of the broad emission line of an AGN could serve as a diagnostic of its binarity. While their method is sensitive to a different part of the binary parameter space (it is only applicable when the binary is more widely separated) and applies emission line (as opposed to broadband) cross-correlation analysis (and thus probes different emitting regions in the system), it similarly raises the possibility that a binary can in principle be identified on a timescale that is significantly shorter than the orbital timescale. 


\begin{acknowledgements}

T.L. thanks the Swift team for scheduling and executing the observations and the referee for helpful comments.
This work is supported by the NANOGrav National Science Foundation Physics Frontiers Center award no.~2020265 and NASA grant 80NSSC24K0251.
Research at UC Irvine was supported by NSF grant AST-1907290. E.M.C. gratefully acknowledges support from the NSF through grant No. AST-1909199.
This work made use of data supplied by the UK Swift Science Data Centre at the University of Leicester and observations from the Las Cumbres Observatory global telescope network.

\end{acknowledgements}

\facilities{Swift, LCO}
\software{\texttt{astropy} \citep{Astropy}; \texttt{matplotlib} \citep{Matplotlib}; \texttt{PyROA} \citep{Donnan2021,Donnan2023}; \texttt{sour} \citep{Edelson2017}}



\begin{thebibliography}{}
\expandafter\ifx\csname natexlab\endcsname\relax\def\natexlab#1{#1}\fi
\providecommand{\url}[1]{\href{#1}{#1}}
\providecommand{\dodoi}[1]{doi:~\href{http://doi.org/#1}{\nolinkurl{#1}}}
\providecommand{\doeprint}[1]{\href{http://ascl.net/#1}{\nolinkurl{http://ascl.net/#1}}}
\providecommand{\doarXiv}[1]{\href{https://arxiv.org/abs/#1}{\nolinkurl{https://arxiv.org/abs/#1}}}

\bibitem[{{Agazie} {et~al.}(2023{\natexlab{a}}){Agazie}, {Anumarlapudi},
  {Archibald}, {Arzoumanian}, {Baker}, {B{\'e}csy}, {Blecha}, {Brazier},
  {Brook}, {Burke-Spolaor}, {Burnette}, {Case}, {Charisi}, {Chatterjee},
  {Chatziioannou}, {Cheeseboro}, {Chen}, {Cohen}, {Cordes}, {Cornish},
  {Crawford}, {Cromartie}, {Crowter}, {Cutler}, {Decesar}, {Degan}, {Demorest},
  {Deng}, {Dolch}, {Drachler}, {Ellis}, {Ferrara}, {Fiore}, {Fonseca},
  {Freedman}, {Garver-Daniels}, {Gentile}, {Gersbach}, {Glaser}, {Good},
  {G{\"u}ltekin}, {Hazboun}, {Hourihane}, {Islo}, {Jennings}, {Johnson},
  {Jones}, {Kaiser}, {Kaplan}, {Kelley}, {Kerr}, {Key}, {Klein}, {Laal}, {Lam},
  {Lamb}, {Lazio}, {Lewandowska}, {Littenberg}, {Liu}, {Lommen}, {Lorimer},
  {Luo}, {Lynch}, {Ma}, {Madison}, {Mattson}, {McEwen}, {McKee}, {McLaughlin},
  {McMann}, {Meyers}, {Meyers}, {Mingarelli}, {Mitridate}, {Natarajan}, {Ng},
  {Nice}, {Ocker}, {Olum}, {Pennucci}, {Perera}, {Petrov}, {Pol}, {Radovan},
  {Ransom}, {Ray}, {Romano}, {Sardesai}, {Schmiedekamp}, {Schmiedekamp},
  {Schmitz}, {Schult}, {Shapiro-Albert}, {Siemens}, {Simon}, {Siwek}, {Stairs},
  {Stinebring}, {Stovall}, {Sun}, {Susobhanan}, {Swiggum}, {Taylor}, {Taylor},
  {Turner}, {Unal}, {Vallisneri}, {van Haasteren}, {Vigeland}, {Wahl}, {Wang},
  {Witt}, {Young}, \& {Nanograv Collaboration}}]{NG15yrGWB}
{Agazie}, G., {Anumarlapudi}, A., {Archibald}, A.~M., {et~al.}
  2023{\natexlab{a}}, \apjl, 951, L8, \dodoi{10.3847/2041-8213/acdac6}

\bibitem[{{Agazie} {et~al.}(2023{\natexlab{b}}){Agazie}, {Anumarlapudi},
  {Archibald}, {Baker}, {B{\'e}csy}, {Blecha}, {Bonilla}, {Brazier}, {Brook},
  {Burke-Spolaor}, {Burnette}, {Case}, {Casey-Clyde}, {Charisi}, {Chatterjee},
  {Chatziioannou}, {Cheeseboro}, {Chen}, {Cohen}, {Cordes}, {Cornish},
  {Crawford}, {Cromartie}, {Crowter}, {Cutler}, {D'Orazio}, {Decesar}, {Degan},
  {Demorest}, {Deng}, {Dolch}, {Drachler}, {Ferrara}, {Fiore}, {Fonseca},
  {Freedman}, {Gardiner}, {Garver-Daniels}, {Gentile}, {Gersbach}, {Glaser},
  {Good}, {G{\"u}ltekin}, {Hazboun}, {Hourihane}, {Islo}, {Jennings},
  {Johnson}, {Jones}, {Kaiser}, {Kaplan}, {Kelley}, {Kerr}, {Key}, {Laal},
  {Lam}, {Lamb}, {Lazio}, {Lewandowska}, {Littenberg}, {Liu}, {Luo}, {Lynch},
  {Ma}, {Madison}, {McEwen}, {McKee}, {McLaughlin}, {McMann}, {Meyers},
  {Meyers}, {Mingarelli}, {Mitridate}, {Natarajan}, {Ng}, {Nice}, {Ocker},
  {Olum}, {Pennucci}, {Perera}, {Petrov}, {Pol}, {Radovan}, {Ransom}, {Ray},
  {Romano}, {Runnoe}, {Sardesai}, {Schmiedekamp}, {Schmiedekamp}, {Schmitz},
  {Schult}, {Shapiro-Albert}, {Siemens}, {Simon}, {Siwek}, {Stairs},
  {Stinebring}, {Stovall}, {Sun}, {Susobhanan}, {Swiggum}, {Taylor}, {Taylor},
  {Turner}, {Unal}, {Vallisneri}, {Vigeland}, {Wachter}, {Wahl}, {Wang},
  {Witt}, {Wright}, {Young}, \& {Nanograv Collaboration}}]{NG15yrAstro}
---. 2023{\natexlab{b}}, \apjl, 952, L37, \dodoi{10.3847/2041-8213/ace18b}

\bibitem[{{Amaro-Seoane} {et~al.}(2017){Amaro-Seoane}, {Audley}, {Babak},
  {Baker}, {Barausse}, {Bender}, {Berti}, {Binetruy}, {Born}, {Bortoluzzi},
  {Camp}, {Caprini}, {Cardoso}, {Colpi}, {Conklin}, {Cornish}, {Cutler},
  {Danzmann}, {Dolesi}, {Ferraioli}, {Ferroni}, {Fitzsimons}, {Gair}, {Gesa
  Bote}, {Giardini}, {Gibert}, {Grimani}, {Halloin}, {Heinzel}, {Hertog},
  {Hewitson}, {Holley-Bockelmann}, {Hollington}, {Hueller}, {Inchauspe},
  {Jetzer}, {Karnesis}, {Killow}, {Klein}, {Klipstein}, {Korsakova}, {Larson},
  {Livas}, {Lloro}, {Man}, {Mance}, {Martino}, {Mateos}, {McKenzie},
  {McWilliams}, {Miller}, {Mueller}, {Nardini}, {Nelemans}, {Nofrarias},
  {Petiteau}, {Pivato}, {Plagnol}, {Porter}, {Reiche}, {Robertson},
  {Robertson}, {Rossi}, {Russano}, {Schutz}, {Sesana}, {Shoemaker}, {Slutsky},
  {Sopuerta}, {Sumner}, {Tamanini}, {Thorpe}, {Troebs}, {Vallisneri},
  {Vecchio}, {Vetrugno}, {Vitale}, {Volonteri}, {Wanner}, {Ward}, {Wass},
  {Weber}, {Ziemer}, \& {Zweifel}}]{Amaro-Seoane2017}
{Amaro-Seoane}, P., {Audley}, H., {Babak}, S., {et~al.} 2017, arXiv e-prints,
  arXiv:1702.00786.
\newblock \doarXiv{1702.00786}

\bibitem[{{Ar{\'e}valo} {et~al.}(2008){Ar{\'e}valo}, {Uttley}, {Kaspi},
  {Breedt}, {Lira}, \& {McHardy}}]{Arevalo2008}
{Ar{\'e}valo}, P., {Uttley}, P., {Kaspi}, S., {et~al.} 2008, \mnras, 389, 1479,
  \dodoi{10.1111/j.1365-2966.2008.13719.x}

\bibitem[{{Artymowicz} \& {Lubow}(1994)}]{Artymowicz1994}
{Artymowicz}, P., \& {Lubow}, S.~H. 1994, \apj, 421, 651,
  \dodoi{10.1086/173679}

\bibitem[{{Astropy Collaboration} {et~al.}(2018){Astropy Collaboration},
  {Price-Whelan}, {Sip{\H{o}}cz}, {G{\"u}nther}, {Lim}, {Crawford}, {Conseil},
  {Shupe}, {Craig}, {Dencheva}, {Ginsburg}, {VanderPlas}, {Bradley},
  {P{\'e}rez-Su{\'a}rez}, {de Val-Borro}, {Aldcroft}, {Cruz}, {Robitaille},
  {Tollerud}, {Ardelean}, {Babej}, {Bach}, {Bachetti}, {Bakanov}, {Bamford},
  {Barentsen}, {Barmby}, {Baumbach}, {Berry}, {Biscani}, {Boquien}, {Bostroem},
  {Bouma}, {Brammer}, {Bray}, {Breytenbach}, {Buddelmeijer}, {Burke},
  {Calderone}, {Cano Rodr{\'\i}guez}, {Cara}, {Cardoso}, {Cheedella}, {Copin},
  {Corrales}, {Crichton}, {D'Avella}, {Deil}, {Depagne}, {Dietrich}, {Donath},
  {Droettboom}, {Earl}, {Erben}, {Fabbro}, {Ferreira}, {Finethy}, {Fox},
  {Garrison}, {Gibbons}, {Goldstein}, {Gommers}, {Greco}, {Greenfield},
  {Groener}, {Grollier}, {Hagen}, {Hirst}, {Homeier}, {Horton}, {Hosseinzadeh},
  {Hu}, {Hunkeler}, {Ivezi{\'c}}, {Jain}, {Jenness}, {Kanarek}, {Kendrew},
  {Kern}, {Kerzendorf}, {Khvalko}, {King}, {Kirkby}, {Kulkarni}, {Kumar},
  {Lee}, {Lenz}, {Littlefair}, {Ma}, {Macleod}, {Mastropietro}, {McCully},
  {Montagnac}, {Morris}, {Mueller}, {Mumford}, {Muna}, {Murphy}, {Nelson},
  {Nguyen}, {Ninan}, {N{\"o}the}, {Ogaz}, {Oh}, {Parejko}, {Parley}, {Pascual},
  {Patil}, {Patil}, {Plunkett}, {Prochaska}, {Rastogi}, {Reddy Janga},
  {Sabater}, {Sakurikar}, {Seifert}, {Sherbert}, {Sherwood-Taylor}, {Shih},
  {Sick}, {Silbiger}, {Singanamalla}, {Singer}, {Sladen}, {Sooley},
  {Sornarajah}, {Streicher}, {Teuben}, {Thomas}, {Tremblay}, {Turner},
  {Terr{\'o}n}, {van Kerkwijk}, {de la Vega}, {Watkins}, {Weaver}, {Whitmore},
  {Woillez}, {Zabalza}, \& {Astropy Contributors}}]{Astropy}
{Astropy Collaboration}, {Price-Whelan}, A.~M., {Sip{\H{o}}cz}, B.~M., {et~al.}
  2018, \aj, 156, 123, \dodoi{10.3847/1538-3881/aabc4f}

\bibitem[{{Begelman} {et~al.}(1980){Begelman}, {Blandford}, \&
  {Rees}}]{Begelman1980}
{Begelman}, M.~C., {Blandford}, R.~D., \& {Rees}, M.~J. 1980, \nat, 287, 307,
  \dodoi{10.1038/287307a0}

\bibitem[{{Blandford} \& {McKee}(1982)}]{Blandford1982}
{Blandford}, R.~D., \& {McKee}, C.~F. 1982, \apj, 255, 419,
  \dodoi{10.1086/159843}

\bibitem[{{Bogdanovi{\'c}} {et~al.}(2022){Bogdanovi{\'c}}, {Miller}, \&
  {Blecha}}]{Bogdanovic2022}
{Bogdanovi{\'c}}, T., {Miller}, M.~C., \& {Blecha}, L. 2022, Living Reviews in
  Relativity, 25, 3, \dodoi{10.1007/s41114-022-00037-8}

\bibitem[{{Brandt} {et~al.}(2018){Brandt}, {Ni}, {Yang}, {Anderson}, {Assef},
  {Barth}, {Bauer}, {Bongiorno}, {Chen}, {De Cicco}, {Gezari}, {Grier}, {Hall},
  {Hoenig}, {Lacy}, {Li}, {Luo}, {Paolillo}, {Peterson}, {Popovi{\'c}},
  {Richards}, {Shemmer}, {Shen}, {Sun}, {Timlin}, {Trump}, {Vito}, \&
  {Yu}}]{Brandt2018}
{Brandt}, W.~N., {Ni}, Q., {Yang}, G., {et~al.} 2018, arXiv e-prints,
  arXiv:1811.06542.
\newblock \doarXiv{1811.06542}

\bibitem[{{Brown} {et~al.}(2013){Brown}, {Baliber}, {Bianco}, {Bowman},
  {Burleson}, {Conway}, {Crellin}, {Depagne}, {De Vera}, {Dilday}, {Dragomir},
  {Dubberley}, {Eastman}, {Elphick}, {Falarski}, {Foale}, {Ford}, {Fulton},
  {Garza}, {Gomez}, {Graham}, {Greene}, {Haldeman}, {Hawkins}, {Haworth},
  {Haynes}, {Hidas}, {Hjelstrom}, {Howell}, {Hygelund}, {Lister}, {Lobdill},
  {Martinez}, {Mullins}, {Norbury}, {Parrent}, {Paulson}, {Petry}, {Pickles},
  {Posner}, {Rosing}, {Ross}, {Sand}, {Saunders}, {Shobbrook}, {Shporer},
  {Street}, {Thomas}, {Tsapras}, {Tufts}, {Valenti}, {Vander Horst}, {Walker},
  {White}, \& {Willis}}]{Brown2013}
{Brown}, T.~M., {Baliber}, N., {Bianco}, F.~B., {et~al.} 2013, \pasp, 125,
  1031, \dodoi{10.1086/673168}

\bibitem[{{Burrows} {et~al.}(2005){Burrows}, {Hill}, {Nousek}, {Kennea},
  {Wells}, {Osborne}, {Abbey}, {Beardmore}, {Mukerjee}, {Short}, {Chincarini},
  {Campana}, {Citterio}, {Moretti}, {Pagani}, {Tagliaferri}, {Giommi},
  {Capalbi}, {Tamburelli}, {Angelini}, {Cusumano}, {Br{\"a}uninger}, {Burkert},
  \& {Hartner}}]{Burrows2005}
{Burrows}, D.~N., {Hill}, J.~E., {Nousek}, J.~A., {et~al.} 2005, \ssr, 120,
  165, \dodoi{10.1007/s11214-005-5097-2}

\bibitem[{{Cackett} {et~al.}(2021){Cackett}, {Bentz}, \& {Kara}}]{Cackett2021}
{Cackett}, E.~M., {Bentz}, M.~C., \& {Kara}, E. 2021, iScience, 24, 102557,
  \dodoi{10.1016/j.isci.2021.102557}

\bibitem[{{Cackett} {et~al.}(2018){Cackett}, {Chiang}, {McHardy}, {Edelson},
  {Goad}, {Horne}, \& {Korista}}]{Cackett2018}
{Cackett}, E.~M., {Chiang}, C.-Y., {McHardy}, I., {et~al.} 2018, \apj, 857, 53,
  \dodoi{10.3847/1538-4357/aab4f7}

\bibitem[{{Cackett} {et~al.}(2007){Cackett}, {Horne}, \&
  {Winkler}}]{Cackett2007}
{Cackett}, E.~M., {Horne}, K., \& {Winkler}, H. 2007, \mnras, 380, 669,
  \dodoi{10.1111/j.1365-2966.2007.12098.x}

\bibitem[{{Cackett} {et~al.}(2020){Cackett}, {Gelbord}, {Li}, {Horne}, {Wang},
  {Barth}, {Bai}, {Bian}, {Carroll}, {Du}, {Edelson}, {Goad}, {Ho}, {Hu},
  {Khatu}, {Luo}, {Miller}, \& {Yuan}}]{Cackett2020}
{Cackett}, E.~M., {Gelbord}, J., {Li}, Y.-R., {et~al.} 2020, \apj, 896, 1,
  \dodoi{10.3847/1538-4357/ab91b5}

\bibitem[{{Carr} {et~al.}(2022){Carr}, {Cinabro}, {Cackett}, {Moutard}, \&
  {Carroll}}]{Carr2022}
{Carr}, R., {Cinabro}, D., {Cackett}, E., {Moutard}, D., \& {Carroll}, R. 2022,
  \pasp, 134, 045002, \dodoi{10.1088/1538-3873/ac5b87}

\bibitem[{{Donnan} {et~al.}(2021){Donnan}, {Horne}, \& {Hern{\'a}ndez
  Santisteban}}]{Donnan2021}
{Donnan}, F.~R., {Horne}, K., \& {Hern{\'a}ndez Santisteban}, J.~V. 2021,
  \mnras, 508, 5449, \dodoi{10.1093/mnras/stab2832}

\bibitem[{{Donnan} {et~al.}(2023){Donnan}, {Hern{\'a}ndez Santisteban},
  {Horne}, {Hu}, {Du}, {Li}, {Xiao}, {Ho}, {Aceituno}, {Wang}, {Guo}, {Yang},
  {Jiang}, \& {Yao}}]{Donnan2023}
{Donnan}, F.~R., {Hern{\'a}ndez Santisteban}, J.~V., {Horne}, K., {et~al.}
  2023, \mnras, 523, 545, \dodoi{10.1093/mnras/stad1409}

\bibitem[{{D'Orazio} {et~al.}(2015){D'Orazio}, {Haiman}, \&
  {Schiminovich}}]{D'Orazio2015}
{D'Orazio}, D.~J., {Haiman}, Z., \& {Schiminovich}, D. 2015, \nat, 525, 351,
  \dodoi{10.1038/nature15262}

\bibitem[{{Dotti} {et~al.}(2023){Dotti}, {Rigamonti}, {Rinaldi}, {Del Pozzo},
  {Decarli}, \& {Buscicchio}}]{Dotti2023}
{Dotti}, M., {Rigamonti}, F., {Rinaldi}, S., {et~al.} 2023, \aap, 680, A69,
  \dodoi{10.1051/0004-6361/202346916}

\bibitem[{{Duffell} {et~al.}(2020){Duffell}, {D'Orazio}, {Derdzinski},
  {Haiman}, {MacFadyen}, {Rosen}, \& {Zrake}}]{Duffell2020}
{Duffell}, P.~C., {D'Orazio}, D., {Derdzinski}, A., {et~al.} 2020, \apj, 901,
  25, \dodoi{10.3847/1538-4357/abab95}

\bibitem[{{Edelson} {et~al.}(2015){Edelson}, {Gelbord}, {Horne}, {McHardy},
  {Peterson}, {Ar{\'e}valo}, {Breeveld}, {De Rosa}, {Evans}, {Goad}, {Kriss},
  {Brandt}, {Gehrels}, {Grupe}, {Kennea}, {Kochanek}, {Nousek}, {Papadakis},
  {Siegel}, {Starkey}, {Uttley}, {Vaughan}, {Young}, {Barth}, {Bentz},
  {Brewer}, {Crenshaw}, {Dalla Bont{\`a}}, {De Lorenzo-C{\'a}ceres}, {Denney},
  {Dietrich}, {Ely}, {Fausnaugh}, {Grier}, {Hall}, {Kaastra}, {Kelly},
  {Korista}, {Lira}, {Mathur}, {Netzer}, {Pancoast}, {Pei}, {Pogge},
  {Schimoia}, {Treu}, {Vestergaard}, {Villforth}, {Yan}, \& {Zu}}]{Edelson2015}
{Edelson}, R., {Gelbord}, J.~M., {Horne}, K., {et~al.} 2015, \apj, 806, 129,
  \dodoi{10.1088/0004-637X/806/1/129}

\bibitem[{{Edelson} {et~al.}(2017){Edelson}, {Gelbord}, {Cackett}, {Connolly},
  {Done}, {Fausnaugh}, {Gardner}, {Gehrels}, {Goad}, {Horne}, {McHardy},
  {Peterson}, {Vaughan}, {Vestergaard}, {Breeveld}, {Barth}, {Bentz},
  {Bottorff}, {Brandt}, {Crawford}, {Dalla Bont{\`a}}, {Emmanoulopoulos},
  {Evans}, {Figuera Jaimes}, {Filippenko}, {Ferland}, {Grupe}, {Joner},
  {Kennea}, {Korista}, {Krimm}, {Kriss}, {Leonard}, {Mathur}, {Netzer},
  {Nousek}, {Page}, {Romero-Colmenero}, {Siegel}, {Starkey}, {Treu}, {Vogler},
  {Winkler}, \& {Zheng}}]{Edelson2017}
{Edelson}, R., {Gelbord}, J., {Cackett}, E., {et~al.} 2017, \apj, 840, 41,
  \dodoi{10.3847/1538-4357/aa6890}

\bibitem[{{Edelson} {et~al.}(2019){Edelson}, {Gelbord}, {Cackett}, {Peterson},
  {Horne}, {Barth}, {Starkey}, {Bentz}, {Brandt}, {Goad}, {Joner}, {Korista},
  {Netzer}, {Page}, {Uttley}, {Vaughan}, {Breeveld}, {Cenko}, {Done}, {Evans},
  {Fausnaugh}, {Ferland}, {Gonzalez-Buitrago}, {Gropp}, {Grupe}, {Kaastra},
  {Kennea}, {Kriss}, {Mathur}, {Mehdipour}, {Mudd}, {Nousek}, {Schmidt},
  {Vestergaard}, \& {Villforth}}]{Edelson2019}
---. 2019, \apj, 870, 123, \dodoi{10.3847/1538-4357/aaf3b4}

\bibitem[{{Evans} {et~al.}(2009){Evans}, {Beardmore}, {Page}, {Osborne},
  {O'Brien}, {Willingale}, {Starling}, {Burrows}, {Godet}, {Vetere}, {Racusin},
  {Goad}, {Wiersema}, {Angelini}, {Capalbi}, {Chincarini}, {Gehrels}, {Kennea},
  {Margutti}, {Morris}, {Mountford}, {Pagani}, {Perri}, {Romano}, \&
  {Tanvir}}]{Evans2009}
{Evans}, P.~A., {Beardmore}, A.~P., {Page}, K.~L., {et~al.} 2009, \mnras, 397,
  1177, \dodoi{10.1111/j.1365-2966.2009.14913.x}

\bibitem[{{Event Horizon Telescope Collaboration} {et~al.}(2019){Event Horizon
  Telescope Collaboration}, {Akiyama}, {Alberdi}, {Alef}, {Asada}, {Azulay},
  {Baczko}, {Ball}, {Balokovi{\'c}}, {Barrett}, {Bintley}, {Blackburn},
  {Boland}, {Bouman}, {Bower}, {Bremer}, {Brinkerink}, {Brissenden}, {Britzen},
  {Broderick}, {Broguiere}, {Bronzwaer}, {Byun}, {Carlstrom}, {Chael}, {Chan},
  {Chatterjee}, {Chatterjee}, {Chen}, {Chen}, {Cho}, {Christian}, {Conway},
  {Cordes}, {Crew}, {Cui}, {Davelaar}, {De Laurentis}, {Deane}, {Dempsey},
  {Desvignes}, {Dexter}, {Doeleman}, {Eatough}, {Falcke}, {Fish}, {Fomalont},
  {Fraga-Encinas}, {Freeman}, {Friberg}, {Fromm}, {G{\'o}mez}, {Galison},
  {Gammie}, {Garc{\'\i}a}, {Gentaz}, {Georgiev}, {Goddi}, {Gold}, {Gu},
  {Gurwell}, {Hada}, {Hecht}, {Hesper}, {Ho}, {Ho}, {Honma}, {Huang}, {Huang},
  {Hughes}, {Ikeda}, {Inoue}, {Issaoun}, {James}, {Jannuzi}, {Janssen},
  {Jeter}, {Jiang}, {Johnson}, {Jorstad}, {Jung}, {Karami}, {Karuppusamy},
  {Kawashima}, {Keating}, {Kettenis}, {Kim}, {Kim}, {Kim}, {Kino}, {Koay},
  {Koch}, {Koyama}, {Kramer}, {Kramer}, {Krichbaum}, {Kuo}, {Lauer}, {Lee},
  {Li}, {Li}, {Lindqvist}, {Liu}, {Liuzzo}, {Lo}, {Lobanov}, {Loinard},
  {Lonsdale}, {Lu}, {MacDonald}, {Mao}, {Markoff}, {Marrone}, {Marscher},
  {Mart{\'\i}-Vidal}, {Matsushita}, {Matthews}, {Medeiros}, {Menten}, {Mizuno},
  {Mizuno}, {Moran}, {Moriyama}, {Moscibrodzka}, {M{\"u}ller}, {Nagai},
  {Nagar}, {Nakamura}, {Narayan}, {Narayanan}, {Natarajan}, {Neri}, {Ni},
  {Noutsos}, {Okino}, {Olivares}, {Ortiz-Le{\'o}n}, {Oyama}, {{\"O}zel},
  {Palumbo}, {Patel}, {Pen}, {Pesce}, {Pi{\'e}tu}, {Plambeck}, {PopStefanija},
  {Porth}, {Prather}, {Preciado-L{\'o}pez}, {Psaltis}, {Pu}, {Ramakrishnan},
  {Rao}, {Rawlings}, {Raymond}, {Rezzolla}, {Ripperda}, {Roelofs}, {Rogers},
  {Ros}, {Rose}, {Roshanineshat}, {Rottmann}, {Roy}, {Ruszczyk}, {Ryan},
  {Rygl}, {S{\'a}nchez}, {S{\'a}nchez-Arguelles}, {Sasada}, {Savolainen},
  {Schloerb}, {Schuster}, {Shao}, {Shen}, {Small}, {Sohn}, {SooHoo}, {Tazaki},
  {Tiede}, {Tilanus}, {Titus}, {Toma}, {Torne}, {Trent}, {Trippe}, {Tsuda},
  {van Bemmel}, {van Langevelde}, {van Rossum}, {Wagner}, {Wardle},
  {Weintroub}, {Wex}, {Wharton}, {Wielgus}, {Wong}, {Wu}, {Young}, {Young},
  {Younsi}, {Yuan}, {Yuan}, {Zensus}, {Zhao}, {Zhao}, {Zhu}, {Algaba},
  {Allardi}, {Amestica}, {Anczarski}, {Bach}, {Baganoff}, {Beaudoin}, {Benson},
  {Berthold}, {Blanchard}, {Blundell}, {Bustamente}, {Cappallo},
  {Castillo-Dom{\'\i}nguez}, {Chang}, {Chang}, {Chang}, {Chen}, {Chilson},
  {Chuter}, {C{\'o}rdova Rosado}, {Coulson}, {Crawford}, {Crowley}, {David},
  {Derome}, {Dexter}, {Dornbusch}, {Dudevoir}, {Dzib}, {Eckart}, {Eckert},
  {Erickson}, {Everett}, {Faber}, {Farah}, {Fath}, {Folkers}, {Forbes},
  {Freund}, {G{\'o}mez-Ruiz}, {Gale}, {Gao}, {Geertsema}, {Graham}, {Greer},
  {Grosslein}, {Gueth}, {Haggard}, {Halverson}, {Han}, {Han}, {Hao},
  {Hasegawa}, {Henning}, {Hern{\'a}ndez-G{\'o}mez}, {Herrero-Illana},
  {Heyminck}, {Hirota}, {Hoge}, {Huang}, {Impellizzeri}, {Jiang}, {Kamble},
  {Keisler}, {Kimura}, {Kono}, {Kubo}, {Kuroda}, {Lacasse}, {Laing}, {Leitch},
  {Li}, {Lin}, {Liu}, {Liu}, {Lu}, {Marson}, {Martin-Cocher}, {Massingill},
  {Matulonis}, {McColl}, {McWhirter}, {Messias}, {Meyer-Zhao}, {Michalik},
  {Monta{\~n}a}, {Montgomerie}, {Mora-Klein}, {Muders}, {Nadolski}, {Navarro},
  {Neilsen}, {Nguyen}, {Nishioka}, {Norton}, {Nowak}, {Nystrom}, {Ogawa},
  {Oshiro}, {Oyama}, {Parsons}, {Paine}, {Pe{\~n}alver}, {Phillips}, {Poirier},
  {Pradel}, {Primiani}, {Raffin}, {Rahlin}, {Reiland}, {Risacher}, {Ruiz},
  {S{\'a}ez-Mada{\'\i}n}, {Sassella}, {Schellart}, {Shaw}, {Silva}, {Shiokawa},
  {Smith}, {Snow}, {Souccar}, {Sousa}, {Sridharan}, {Srinivasan}, {Stahm},
  {Stark}, {Story}, {Timmer}, {Vertatschitsch}, {Walther}, {Wei}, {Whitehorn},
  {Whitney}, {Woody}, {Wouterloot}, {Wright}, {Yamaguchi}, {Yu}, {Zeballos},
  {Zhang}, \& {Ziurys}}]{EHT2019}
{Event Horizon Telescope Collaboration}, {Akiyama}, K., {Alberdi}, A., {et~al.}
  2019, \apjl, 875, L1, \dodoi{10.3847/2041-8213/ab0ec7}

\bibitem[{{Event Horizon Telescope Collaboration} {et~al.}(2022){Event Horizon
  Telescope Collaboration}, {Akiyama}, {Alberdi}, {Alef}, {Algaba}, {Anantua},
  {Asada}, {Azulay}, {Bach}, {Baczko}, {Ball}, {Balokovi{\'c}}, {Barrett},
  {Baub{\"o}ck}, {Benson}, {Bintley}, {Blackburn}, {Blundell}, {Bouman},
  {Bower}, {Boyce}, {Bremer}, {Brinkerink}, {Brissenden}, {Britzen},
  {Broderick}, {Broguiere}, {Bronzwaer}, {Bustamante}, {Byun}, {Carlstrom},
  {Ceccobello}, {Chael}, {Chan}, {Chatterjee}, {Chatterjee}, {Chen}, {Chen},
  {Cheng}, {Cho}, {Christian}, {Conroy}, {Conway}, {Cordes}, {Crawford},
  {Crew}, {Cruz-Osorio}, {Cui}, {Davelaar}, {De Laurentis}, {Deane}, {Dempsey},
  {Desvignes}, {Dexter}, {Dhruv}, {Doeleman}, {Dougal}, {Dzib}, {Eatough},
  {Emami}, {Falcke}, {Farah}, {Fish}, {Fomalont}, {Ford}, {Fraga-Encinas},
  {Freeman}, {Friberg}, {Fromm}, {Fuentes}, {Galison}, {Gammie}, {Garc{\'\i}a},
  {Gentaz}, {Georgiev}, {Goddi}, {Gold}, {G{\'o}mez-Ruiz}, {G{\'o}mez}, {Gu},
  {Gurwell}, {Hada}, {Haggard}, {Haworth}, {Hecht}, {Hesper}, {Heumann}, {Ho},
  {Ho}, {Honma}, {Huang}, {Huang}, {Hughes}, {Ikeda}, {Impellizzeri}, {Inoue},
  {Issaoun}, {James}, {Jannuzi}, {Janssen}, {Jeter}, {Jiang},
  {Jim{\'e}nez-Rosales}, {Johnson}, {Jorstad}, {Joshi}, {Jung}, {Karami},
  {Karuppusamy}, {Kawashima}, {Keating}, {Kettenis}, {Kim}, {Kim}, {Kim},
  {Kim}, {Kino}, {Koay}, {Kocherlakota}, {Kofuji}, {Koch}, {Koyama}, {Kramer},
  {Kramer}, {Krichbaum}, {Kuo}, {La Bella}, {Lauer}, {Lee}, {Lee}, {Leung},
  {Levis}, {Li}, {Lico}, {Lindahl}, {Lindqvist}, {Lisakov}, {Liu}, {Liu},
  {Liuzzo}, {Lo}, {Lobanov}, {Loinard}, {Lonsdale}, {Lu}, {Mao}, {Marchili},
  {Markoff}, {Marrone}, {Marscher}, {Mart{\'\i}-Vidal}, {Matsushita},
  {Matthews}, {Medeiros}, {Menten}, {Michalik}, {Mizuno}, {Mizuno}, {Moran},
  {Moriyama}, {Moscibrodzka}, {M{\"u}ller}, {Mus}, {Musoke}, {Myserlis},
  {Nadolski}, {Nagai}, {Nagar}, {Nakamura}, {Narayan}, {Narayanan},
  {Natarajan}, {Nathanail}, {Fuentes}, {Neilsen}, {Neri}, {Ni}, {Noutsos},
  {Nowak}, {Oh}, {Okino}, {Olivares}, {Ortiz-Le{\'o}n}, {Oyama}, {{\"O}zel},
  {Palumbo}, {Paraschos}, {Park}, {Parsons}, {Patel}, {Pen}, {Pesce},
  {Pi{\'e}tu}, {Plambeck}, {PopStefanija}, {Porth}, {P{\"o}tzl}, {Prather},
  {Preciado-L{\'o}pez}, {Psaltis}, {Pu}, {Ramakrishnan}, {Rao}, {Rawlings},
  {Raymond}, {Rezzolla}, {Ricarte}, {Ripperda}, {Roelofs}, {Rogers}, {Ros},
  {Romero-Ca{\~n}izales}, {Roshanineshat}, {Rottmann}, {Roy}, {Ruiz},
  {Ruszczyk}, {Rygl}, {S{\'a}nchez}, {S{\'a}nchez-Arg{\"u}elles},
  {S{\'a}nchez-Portal}, {Sasada}, {Satapathy}, {Savolainen}, {Schloerb},
  {Schonfeld}, {Schuster}, {Shao}, {Shen}, {Small}, {Sohn}, {SooHoo},
  {Souccar}, {Sun}, {Tazaki}, {Tetarenko}, {Tiede}, {Tilanus}, {Titus},
  {Torne}, {Traianou}, {Trent}, {Trippe}, {Turk}, {van Bemmel}, {van
  Langevelde}, {van Rossum}, {Vos}, {Wagner}, {Ward-Thompson}, {Wardle},
  {Weintroub}, {Wex}, {Wharton}, {Wielgus}, {Wiik}, {Witzel}, {Wondrak},
  {Wong}, {Wu}, {Yamaguchi}, {Yoon}, {Young}, {Young}, {Younsi}, {Yuan},
  {Yuan}, {Zensus}, {Zhang}, {Zhao}, {Zhao}, {Agurto}, {Allardi}, {Amestica},
  {Araneda}, {Arriagada}, {Berghuis}, {Bertarini}, {Berthold}, {Blanchard},
  {Brown}, {C{\'a}rdenas}, {Cantzler}, {Caro}, {Castillo-Dom{\'\i}nguez},
  {Chan}, {Chang}, {Chang}, {Chang}, {Chang}, {Chen}, {Chilson}, {Chuter},
  {Ciechanowicz}, {Colin-Beltran}, {Coulson}, {Crowley}, {Degenaar},
  {Dornbusch}, {Dur{\'a}n}, {Everett}, {Faber}, {Forster}, {Fuchs}, {Gale},
  {Geertsema}, {Gonz{\'a}lez}, {Graham}, {Gueth}, {Halverson}, {Han}, {Han},
  {Hasegawa}, {Hern{\'a}ndez-Rebollar}, {Herrera}, {Herrero-Illana},
  {Heyminck}, {Hirota}, {Hoge}, {Hostler Schimpf}, {Howie}, {Huang}, {Jiang},
  {Jinchi}, {John}, {Kimura}, {Klein}, {Kubo}, {Kuroda}, {Kwon}, {Lacasse},
  {Laing}, {Leitch}, {Li}, {Liu}, {Liu}, {Lin}, {Lu}, {Mac-Auliffe},
  {Martin-Cocher}, {Matulonis}, {Maute}, {Messias}, {Meyer-Zhao},
  {Monta{\~n}a}, {Montenegro-Montes}, {Montgomerie}, {Moreno Nolasco},
  {Muders}, {Nishioka}, {Norton}, {Nystrom}, {Ogawa}, {Olivares}, {Oshiro},
  {P{\'e}rez-Beaupuits}, {Parra}, {Phillips}, {Poirier}, {Pradel}, {Qiu},
  {Raffin}, {Rahlin}, {Ram{\'\i}rez}, {Ressler}, {Reynolds},
  {Rodr{\'\i}guez-Montoya}, {Saez-Madain}, {Santana}, {Shaw}, {Shirkey},
  {Silva}, {Snow}, {Sousa}, {Sridharan}, {Stahm}, {Stark}, {Test},
  {Torstensson}, {Venegas}, {Walther}, {Wei}, {White}, {Wieching}, {Wijnands},
  {Wouterloot}, {Yu}, {Yu (于威)}, \& {Zeballos}}]{EHT2022}
---. 2022, \apjl, 930, L12, \dodoi{10.3847/2041-8213/ac6674}

\bibitem[{{Farris} {et~al.}(2014){Farris}, {Duffell}, {MacFadyen}, \&
  {Haiman}}]{Farris2014}
{Farris}, B.~D., {Duffell}, P., {MacFadyen}, A.~I., \& {Haiman}, Z. 2014, ApJ,
  783, 134, \dodoi{10.1088/0004-637X/783/2/134}

\bibitem[{{Fausnaugh} {et~al.}(2016){Fausnaugh}, {Denney}, {Barth}, {Bentz},
  {Bottorff}, {Carini}, {Croxall}, {De Rosa}, {Goad}, {Horne}, {Joner},
  {Kaspi}, {Kim}, {Klimanov}, {Kochanek}, {Leonard}, {Netzer}, {Peterson},
  {Schn{\"u}lle}, {Sergeev}, {Vestergaard}, {Zheng}, {Zu}, {Anderson},
  {Ar{\'e}valo}, {Bazhaw}, {Borman}, {Boroson}, {Brandt}, {Breeveld}, {Brewer},
  {Cackett}, {Crenshaw}, {Dalla Bont{\`a}}, {De Lorenzo-C{\'a}ceres},
  {Dietrich}, {Edelson}, {Efimova}, {Ely}, {Evans}, {Filippenko}, {Flatland},
  {Gehrels}, {Geier}, {Gelbord}, {Gonzalez}, {Gorjian}, {Grier}, {Grupe},
  {Hall}, {Hicks}, {Horenstein}, {Hutchison}, {Im}, {Jensen}, {Jones},
  {Kaastra}, {Kelly}, {Kennea}, {Kim}, {Korista}, {Kriss}, {Lee}, {Lira},
  {MacInnis}, {Manne-Nicholas}, {Mathur}, {McHardy}, {Montouri}, {Musso},
  {Nazarov}, {Norris}, {Nousek}, {Okhmat}, {Pancoast}, {Papadakis}, {Parks},
  {Pei}, {Pogge}, {Pott}, {Rafter}, {Rix}, {Saylor}, {Schimoia}, {Siegel},
  {Spencer}, {Starkey}, {Sung}, {Teems}, {Treu}, {Turner}, {Uttley},
  {Villforth}, {Weiss}, {Woo}, {Yan}, \& {Young}}]{Fausnaugh2016}
{Fausnaugh}, M.~M., {Denney}, K.~D., {Barth}, A.~J., {et~al.} 2016, \apj, 821,
  56, \dodoi{10.3847/0004-637X/821/1/56}

\bibitem[{{Gaskell} \& {Sparke}(1986)}]{Gaskell1986}
{Gaskell}, C.~M., \& {Sparke}, L.~S. 1986, \apj, 305, 175,
  \dodoi{10.1086/164238}

\bibitem[{{Gehrels} {et~al.}(2004){Gehrels}, {Chincarini}, {Giommi}, {Mason},
  {Nousek}, {Wells}, {White}, {Barthelmy}, {Burrows}, {Cominsky}, {Hurley},
  {Marshall}, {M{\'e}sz{\'a}ros}, {Roming}, {Angelini}, {Barbier}, {Belloni},
  {Campana}, {Caraveo}, {Chester}, {Citterio}, {Cline}, {Cropper}, {Cummings},
  {Dean}, {Feigelson}, {Fenimore}, {Frail}, {Fruchter}, {Garmire}, {Gendreau},
  {Ghisellini}, {Greiner}, {Hill}, {Hunsberger}, {Krimm}, {Kulkarni}, {Kumar},
  {Lebrun}, {Lloyd-Ronning}, {Markwardt}, {Mattson}, {Mushotzky}, {Norris},
  {Osborne}, {Paczynski}, {Palmer}, {Park}, {Parsons}, {Paul}, {Rees},
  {Reynolds}, {Rhoads}, {Sasseen}, {Schaefer}, {Short}, {Smale}, {Smith},
  {Stella}, {Tagliaferri}, {Takahashi}, {Tashiro}, {Townsley}, {Tueller},
  {Turner}, {Vietri}, {Voges}, {Ward}, {Willingale}, {Zerbi}, \&
  {Zhang}}]{Gehrels2004}
{Gehrels}, N., {Chincarini}, G., {Giommi}, P., {et~al.} 2004, \apj, 611, 1005,
  \dodoi{10.1086/422091}

\bibitem[{{Graham} {et~al.}(2015){Graham}, {Djorgovski}, {Stern}, {Glikman},
  {Drake}, {Mahabal}, {Donalek}, {Larson}, \& {Christensen}}]{Graham2015Nat}
{Graham}, M.~J., {Djorgovski}, S.~G., {Stern}, D., {et~al.} 2015, \nat, 518,
  74, \dodoi{10.1038/nature14143}

\bibitem[{{Guo} {et~al.}(2020){Guo}, {Liu}, {Zafar}, \& {Liao}}]{Guo2020}
{Guo}, H., {Liu}, X., {Zafar}, T., \& {Liao}, W.-T. 2020, \mnras, 492, 2910,
  \dodoi{10.1093/mnras/stz3566}

\bibitem[{{Hern{\'a}ndez Santisteban} {et~al.}(2020){Hern{\'a}ndez
  Santisteban}, {Edelson}, {Horne}, {Gelbord}, {Barth}, {Cackett}, {Goad},
  {Netzer}, {Starkey}, {Uttley}, {Brandt}, {Korista}, {Lohfink}, {Onken},
  {Page}, {Siegel}, {Vestergaard}, {Bisogni}, {Breeveld}, {Cenko}, {Dalla
  Bont{\`a}}, {Evans}, {Ferland}, {Gonzalez-Buitrago}, {Grupe}, {Joner},
  {Kriss}, {LaPorte}, {Mathur}, {Marshall}, {Mehdipour}, {Mudd}, {Peterson},
  {Schmidt}, {Vaughan}, \& {Valenti}}]{Hernandez2020}
{Hern{\'a}ndez Santisteban}, J.~V., {Edelson}, R., {Horne}, K., {et~al.} 2020,
  \mnras, 498, 5399, \dodoi{10.1093/mnras/staa2365}

\bibitem[{Hunter(2007)}]{Matplotlib}
Hunter, J.~D. 2007, Computing in Science \& Engineering, 9, 90,
  \dodoi{10.1109/MCSE.2007.55}

\bibitem[{{Ivezic} {et~al.}(2008){Ivezic}, {Tyson}, {Abel}, {Acosta},
  {Allsman}, {AlSayyad}, {Anderson}, {Andrew}, {Angel}, {Angeli}, {Ansari},
  {Antilogus}, {Arndt}, {Astier}, {Aubourg}, {Axelrod}, {Bard}, {Barr},
  {Barrau}, {Bartlett}, {Bauman}, {Beaumont}, {Becker}, {Becla}, {Beldica},
  {Bellavia}, {Blanc}, {Blandford}, {Bloom}, {Bogart}, {Borne}, {Bosch},
  {Boutigny}, {Brandt}, {Brown}, {Bullock}, {Burchat}, {Burke}, {Cagnoli},
  {Calabrese}, {Chandrasekharan}, {Chesley}, {Cheu}, {Chiang}, {Claver},
  {Connolly}, {Cook}, {Cooray}, {Covey}, {Cribbs}, {Cui}, {Cutri}, {Daubard},
  {Daues}, {Delgado}, {Digel}, {Doherty}, {Dubois}, {Dubois-Felsmann},
  {Durech}, {Eracleous}, {Ferguson}, {Frank}, {Freemon}, {Gangler}, {Gawiser},
  {Geary}, {Gee}, {Geha}, {Gibson}, {Gilmore}, {Glanzman}, {Goodenow},
  {Gressler}, {Gris}, {Guyonnet}, {Hascall}, {Haupt}, {Hernandez}, {Hogan},
  {Huang}, {Huffer}, {Innes}, {Jacoby}, {Jain}, {Jee}, {Jernigan},
  {Jevremovic}, {Johns}, {Jones}, {Juramy-Gilles}, {Juric}, {Kahn}, {Kalirai},
  {Kallivayalil}, {Kalmbach}, {Kantor}, {Kasliwal}, {Kessler}, {Kirkby},
  {Knox}, {Kotov}, {Krabbendam}, {Krughoff}, {Kubanek}, {Kuczewski},
  {Kulkarni}, {Lambert}, {Le Guillou}, {Levine}, {Liang}, {Lim}, {Lintott},
  {Lupton}, {Mahabal}, {Marshall}, {Marshall}, {May}, {McKercher}, {Migliore},
  {Miller}, {Mills}, {Monet}, {Moniez}, {Neill}, {Nief}, {Nomerotski},
  {Nordby}, {O'Connor}, {Oliver}, {Olivier}, {Olsen}, {Ortiz}, {Owen}, {Pain},
  {Peterson}, {Petry}, {Pierfederici}, {Pietrowicz}, {Pike}, {Pinto}, {Plante},
  {Plate}, {Price}, {Prouza}, {Radeka}, {Rajagopal}, {Rasmussen}, {Regnault},
  {Ridgway}, {Ritz}, {Rosing}, {Roucelle}, {Rumore}, {Russo}, {Saha},
  {Sassolas}, {Schalk}, {Schindler}, {Schneider}, {Schumacher}, {Sebag},
  {Sembroski}, {Seppala}, {Shipsey}, {Silvestri}, {Smith}, {Smith}, {Strauss},
  {Stubbs}, {Sweeney}, {Szalay}, {Takacs}, {Thaler}, {Van Berg}, {Vanden Berk},
  {Vetter}, {Virieux}, {Xin}, {Walkowicz}, {Walter}, {Wang}, {Warner},
  {Willman}, {Wittman}, {Wolff}, {Wood-Vasey}, {Yoachim}, {Zhan}, \& {for the
  LSST Collaboration}}]{Ivezic2008}
{Ivezic}, Z., {Tyson}, J.~A., {Abel}, B., {et~al.} 2008, ArXiv e-prints.
\newblock \doarXiv{0805.2366}

\bibitem[{{Kara} {et~al.}(2021){Kara}, {Mehdipour}, {Kriss}, {Cackett}, {Arav},
  {Barth}, {Byun}, {Brotherton}, {De Rosa}, {Gelbord}, {Hern{\'a}ndez
  Santisteban}, {Hu}, {Kaastra}, {Landt}, {Li}, {Miller}, {Montano},
  {Partington}, {Aceituno}, {Bai}, {Bao}, {Bentz}, {Brink}, {Chelouche},
  {Chen}, {Colmenero}, {Dalla Bont{\`a}}, {Dehghanian}, {Du}, {Edelson},
  {Ferland}, {Ferrarese}, {Fian}, {Filippenko}, {Fischer}, {Goad},
  {Gonz{\'a}lez Buitrago}, {Gorjian}, {Grier}, {Guo}, {Hall}, {Ho},
  {Homayouni}, {Horne}, {Ili{\'c}}, {Jiang}, {Joner}, {Kaspi}, {Kochanek},
  {Korista}, {Kynoch}, {Li}, {Liu}, {McHardy}, {McLane}, {Mitchell}, {Netzer},
  {Olson}, {Pogge}, {Popovi{\'c}}, {Proga}, {Storchi-Bergmann}, {Strasburger},
  {Treu}, {Vestergaard}, {Wang}, {Ward}, {Waters}, {Williams}, {Yang}, {Yao},
  {Zastrocky}, {Zhai}, \& {Zu}}]{Kara2021}
{Kara}, E., {Mehdipour}, M., {Kriss}, G.~A., {et~al.} 2021, \apj, 922, 151,
  \dodoi{10.3847/1538-4357/ac2159}

\bibitem[{{Kara} {et~al.}(2023){Kara}, {Barth}, {Cackett}, {Gelbord},
  {Montano}, {Li}, {Santana}, {Horne}, {Alston}, {Buisson}, {Chelouche}, {Du},
  {Fabian}, {Fian}, {Gallo}, {Goad}, {Grupe}, {Gonz{\'a}lez Buitrago},
  {Hern{\'a}ndez Santisteban}, {Kaspi}, {Hu}, {Komossa}, {Kriss}, {Lewin},
  {Lewis}, {Loewenstein}, {Lohfink}, {Masterson}, {McHardy}, {Mehdipour},
  {Miller}, {Panagiotou}, {Parker}, {Pinto}, {Remillard}, {Reynolds},
  {Rogantini}, {Wang}, {Wang}, \& {Wilkins}}]{Kara2023}
{Kara}, E., {Barth}, A.~J., {Cackett}, E.~M., {et~al.} 2023, \apj, 947, 62,
  \dodoi{10.3847/1538-4357/acbcd3}

\bibitem[{{Komossa} {et~al.}(2021){Komossa}, {Grupe}, {Kraus}, {Gallo},
  {Gonzalez}, {Parker}, {Valtonen}, {Hollett}, {Bach}, {G{\'o}mez}, {Myserlis},
  \& {Ciprini}}]{Komossa2021MOMO}
{Komossa}, S., {Grupe}, D., {Kraus}, A., {et~al.} 2021, Universe, 7, 261,
  \dodoi{10.3390/universe7080261}

\bibitem[{{Korista} \& {Goad}(2019)}]{Korista2019}
{Korista}, K.~T., \& {Goad}, M.~R. 2019, \mnras, 489, 5284,
  \dodoi{10.1093/mnras/stz2330}

\bibitem[{{Li} {et~al.}(2014){Li}, {Wang}, {Hu}, {Du}, \& {Bai}}]{Li2014}
{Li}, Y.-R., {Wang}, J.-M., {Hu}, C., {Du}, P., \& {Bai}, J.-M. 2014, \apjl,
  786, L6, \dodoi{10.1088/2041-8205/786/1/L6}

\bibitem[{{Liu} {et~al.}(2018){Liu}, {Gezari}, \& {Miller}}]{Liu2018}
{Liu}, T., {Gezari}, S., \& {Miller}, M.~C. 2018, \apjl, 859, L12,
  \dodoi{10.3847/2041-8213/aac2ed}

\bibitem[{{McCully} {et~al.}(2018){McCully}, {Volgenau}, {Harbeck}, {Lister},
  {Saunders}, {Turner}, {Siiverd}, \& {Bowman}}]{McCully2018}
{McCully}, C., {Volgenau}, N.~H., {Harbeck}, D.-R., {et~al.} 2018, in Society
  of Photo-Optical Instrumentation Engineers (SPIE) Conference Series, Vol.
  10707, Software and Cyberinfrastructure for Astronomy V, ed. J.~C. {Guzman}
  \& J.~{Ibsen}, 107070K, \dodoi{10.1117/12.2314340}

\bibitem[{{McHardy} {et~al.}(2014){McHardy}, {Cameron}, {Dwelly}, {Connolly},
  {Lira}, {Emmanoulopoulos}, {Gelbord}, {Breedt}, {Arevalo}, \&
  {Uttley}}]{McHardy2014}
{McHardy}, I.~M., {Cameron}, D.~T., {Dwelly}, T., {et~al.} 2014, \mnras, 444,
  1469, \dodoi{10.1093/mnras/stu1636}

\bibitem[{{McHardy} {et~al.}(2016){McHardy}, {Connolly}, {Peterson}, {Bieryla},
  {Chand}, {Elvis}, {Emmanoulopoulos}, {Falco}, {Gandhi}, {Kaspi}, {Latham},
  {Lira}, {McCully}, {Netzer}, \& {Uemura}}]{McHardy2016}
{McHardy}, I.~M., {Connolly}, S.~D., {Peterson}, B.~M., {et~al.} 2016,
  Astronomische Nachrichten, 337, 500, \dodoi{10.1002/asna.201612337}

\bibitem[{{McHardy} {et~al.}(2018){McHardy}, {Connolly}, {Horne}, {Cackett},
  {Gelbord}, {Peterson}, {Pahari}, {Gehrels}, {Goad}, {Lira}, {Arevalo},
  {Baldi}, {Brandt}, {Breedt}, {Chand}, {Dewangan}, {Done}, {Elvis},
  {Emmanoulopoulos}, {Fausnaugh}, {Kaspi}, {Kochanek}, {Korista}, {Papadakis},
  {Rao}, {Uttley}, {Vestergaard}, \& {Ward}}]{McHardy2018}
{McHardy}, I.~M., {Connolly}, S.~D., {Horne}, K., {et~al.} 2018, \mnras, 480,
  2881, \dodoi{10.1093/mnras/sty1983}

\bibitem[{{Peterson}(1993)}]{Peterson1993}
{Peterson}, B.~M. 1993, \pasp, 105, 247, \dodoi{10.1086/133140}

\bibitem[{{Peterson} {et~al.}(1998){Peterson}, {Wanders}, {Horne}, {Collier},
  {Alexander}, {Kaspi}, \& {Maoz}}]{Peterson1998}
{Peterson}, B.~M., {Wanders}, I., {Horne}, K., {et~al.} 1998, \pasp, 110, 660,
  \dodoi{10.1086/316177}

\bibitem[{{Rodriguez} {et~al.}(2006){Rodriguez}, {Taylor}, {Zavala}, {Peck},
  {Pollack}, \& {Romani}}]{Rodriguez2006}
{Rodriguez}, C., {Taylor}, G.~B., {Zavala}, R.~T., {et~al.} 2006, ApJ, 646, 49,
  \dodoi{10.1086/504825}

\bibitem[{{Roedig} {et~al.}(2014){Roedig}, {Krolik}, \& {Miller}}]{Roedig2014}
{Roedig}, C., {Krolik}, J.~H., \& {Miller}, M.~C. 2014, \apj, 785, 115,
  \dodoi{10.1088/0004-637X/785/2/115}

\bibitem[{{Roming} {et~al.}(2005){Roming}, {Kennedy}, {Mason}, {Nousek}, {Ahr},
  {Bingham}, {Broos}, {Carter}, {Hancock}, {Huckle}, {Hunsberger}, {Kawakami},
  {Killough}, {Koch}, {McLelland}, {Smith}, {Smith}, {Soto}, {Boyd},
  {Breeveld}, {Holland}, {Ivanushkina}, {Pryzby}, {Still}, \&
  {Stock}}]{Roming2005}
{Roming}, P. W.~A., {Kennedy}, T.~E., {Mason}, K.~O., {et~al.} 2005, \ssr, 120,
  95, \dodoi{10.1007/s11214-005-5095-4}

\bibitem[{{Shappee} {et~al.}(2014){Shappee}, {Prieto}, {Grupe}, {Kochanek},
  {Stanek}, {De Rosa}, {Mathur}, {Zu}, {Peterson}, {Pogge}, {Komossa}, {Im},
  {Jencson}, {Holoien}, {Basu}, {Beacom}, {Szczygie{\l}}, {Brimacombe},
  {Adams}, {Campillay}, {Choi}, {Contreras}, {Dietrich}, {Dubberley},
  {Elphick}, {Foale}, {Giustini}, {Gonzalez}, {Hawkins}, {Howell}, {Hsiao},
  {Koss}, {Leighly}, {Morrell}, {Mudd}, {Mullins}, {Nugent}, {Parrent},
  {Phillips}, {Pojmanski}, {Rosing}, {Ross}, {Sand}, {Terndrup}, {Valenti},
  {Walker}, \& {Yoon}}]{Shappee2014}
{Shappee}, B.~J., {Prieto}, J.~L., {Grupe}, D., {et~al.} 2014, \apj, 788, 48,
  \dodoi{10.1088/0004-637X/788/1/48}

\bibitem[{{Valtonen} {et~al.}(2021){Valtonen}, {Dey}, {Gopakumar}, {Zola},
  {Komossa}, {Pursimo}, {Gomez}, {Hudec}, {Jermak}, \&
  {Berdyugin}}]{Valtonen2021}
{Valtonen}, M.~J., {Dey}, L., {Gopakumar}, A., {et~al.} 2021, Galaxies, 10, 1,
  \dodoi{10.3390/galaxies10010001}

\bibitem[{{Vaughan} {et~al.}(2003){Vaughan}, {Edelson}, {Warwick}, \&
  {Uttley}}]{Vaughan2003}
{Vaughan}, S., {Edelson}, R., {Warwick}, R.~S., \& {Uttley}, P. 2003, \mnras,
  345, 1271, \dodoi{10.1046/j.1365-2966.2003.07042.x}

\bibitem[{{Vaughan} {et~al.}(2016){Vaughan}, {Uttley}, {Markowitz},
  {Huppenkothen}, {Middleton}, {Alston}, {Scargle}, \& {Farr}}]{Vaughan2016}
{Vaughan}, S., {Uttley}, P., {Markowitz}, A.~G., {et~al.} 2016, \mnras, 461,
  3145, \dodoi{10.1093/mnras/stw1412}

\bibitem[{{Vincentelli} {et~al.}(2022){Vincentelli}, {McHardy}, {Hern{\'a}ndez
  Santisteban}, {Cackett}, {Gelbord}, {Horne}, {Miller}, \&
  {Lobban}}]{Vincentelli2022}
{Vincentelli}, F.~M., {McHardy}, I., {Hern{\'a}ndez Santisteban}, J.~V.,
  {et~al.} 2022, \mnras, 512, L33, \dodoi{10.1093/mnrasl/slac009}

\bibitem[{{Vincentelli} {et~al.}(2021){Vincentelli}, {McHardy}, {Cackett},
  {Barth}, {Horne}, {Goad}, {Korista}, {Gelbord}, {Brandt}, {Edelson},
  {Miller}, {Pahari}, {Peterson}, {Schmidt}, {Baldi}, {Breedt}, {Hern{\'a}ndez
  Santisteban}, {Romero-Colmenero}, {Ward}, \& {Williams}}]{Vincentelli2021}
{Vincentelli}, F.~M., {McHardy}, I., {Cackett}, E.~M., {et~al.} 2021, \mnras,
  504, 4337, \dodoi{10.1093/mnras/stab1033}

\bibitem[{{Yu} \& {Tremaine}(2002)}]{Yu2002}
{Yu}, Q., \& {Tremaine}, S. 2002, \mnras, 335, 965,
  \dodoi{10.1046/j.1365-8711.2002.05532.x}

\end{thebibliography}

\end{document}